\documentclass[a4paper,fleqn,usenatbib]{mnras}

\usepackage{newtxtext,newtxmath}
\usepackage[T1]{fontenc}
\usepackage{ae,aecompl}
\usepackage{graphicx}	% Including figure files
\usepackage{amsmath}	% Advanced maths commands
\usepackage{amssymb}	% Extra maths symbols
\usepackage{topcapt}
\usepackage{supertabular}
\usepackage{adjustbox}
\usepackage{hyperref}
\usepackage{stfloats}
\hypersetup{colorlinks}

\title[Automatic Strong Lens-Modeling of Massive Clusters]{AStroLens: Automatic Strong-Lens Modeling of X-ray Selected Galaxy Clusters}
\author[L. Zalesky \& H. Ebeling]{
Lukas Zalesky,$^{1}$\thanks{E-mail: zalesky@hawaii.edu}
Harald Ebeling$^{1}$ 
\\
$^{1}$Institute for Astronomy University of Hawaii, 2680 Woodlawn Drive Honolulu, HI 96822, USA 
}
\date{Accepted 2020 July 22. Received 2020 July 21; in original form 2020 May 9.}
\pubyear{2020}
\begin{document}
\renewcommand{\floatpagefraction}{1}

\label{firstpage}
\pagerange{\pageref{firstpage}--\pageref{lastpage}}
\maketitle

%===================================================================================================
\begin{abstract}
We use \texttt{AStroLens}, a newly developed gravitational lens-modeling code that relies only on geometric and photometric information of cluster galaxies as input, to map the strong-lensing regions and estimate the lensing strength of 96 galaxy clusters at $z=0.5$--$0.9$. All clusters were identified during the extended Massive Cluster Survey (eMACS) based on their X-ray flux and optical appearance. Building on the well tested assumption that the distribution of both luminous and dark matter in galaxy clusters is approximately traced by the distribution of light, i.e., that light traces mass, \texttt{AStroLens} uses three global parameters to automatically model the deflection from strong-gravitational lensing for all galaxy clusters in this diverse sample. We test the robustness of our code by comparing \texttt{AStroLens} estimates derived solely from shallow optical images in two passbands with the results of in-depth lens-modeling efforts for two well studied eMACS clusters and find good agreement, both with respect to the size and the shape of the strong-lensing regime delineated by the respective critical lines. Our study finds 31 eMACS clusters with effective Einstein radii ($\theta_{\rm E}$) in excess of $20"$ and eight with $\theta_{\rm E} > 30"$, thereby underlining the value of X-ray selection for the discovery of powerful cluster lenses that complement giants like MACSJ0717 at ever-increasing redshift. As a first installment toward the public release of the eMACS sample, we list physical properties of the ten calibration clusters as well as of the ten most powerful eMACS cluster lenses, according to \texttt{AStroLens}.

\end{abstract}


%===================================================================================================
\section{Introduction}
As the most massive bound structures in the Universe, galaxy clusters constitute the most powerful gravitational lenses known, opening a window into the high-redshift Universe and providing a tool to map and understand the unseen, yet dominant component of all gravitational mass in the Universe, dark matter (DM). High surface-mass density is needed for strong-gravitational lensing to occur, a requirement that is met within the immediate vicinity of many astronomical objects, from stars to individual galaxies. However, only galaxy clusters, having evolved from the highest peaks of the primordial density field, are massive enough to enable strong gravitational lensing across significant areas (up to a square arcminute, ~\citealt{2011A&ARv..19...47K}).

When the surface mass density of the cluster lens exceeds a critical threshold, multiple images of the same source can be seen at different locations in the plane of the sky with varying magnifications. The diversity of research made possible by studies of strong-lensing clusters, specifically in the multiple-image regime, has motivated several large surveys, including the Cluster Lensing and Supernova Survey with Hubble (CLASH; ~\citealt{2012ApJS..199...25P}), the Hubble Frontier Fields (HFF; ~\citealt{2017ApJ...837...97L}), and most recently, the Reionization Lensing Cluster Survey (RELICS; ~\citealt{2018ApJ...859..159C}). Observations of powerful cluster lenses have enabled both detailed studies of representative galaxies at intermediate redshift \citep[e.g.,][]{2007ApJ...654L..33S,2008Natur.455..775S,2009MNRAS.398.1263Q,2011ApJ...742...11S,2011ApJ...732L..14Y,2014ApJ...790..144B,2015ApJ...813L...7N,2016ApJ...817...60T,2017Natur.546..510T,2017MNRAS.467.3306S,2018ApJ...852L...7E} and statistical investigations of galaxy evolution in the early universe \citep[e.g.,][]{1996ApJ...471..643K,2012Natur.489..406Z,2014ApJ...795..126B,2015ApJ...799...12I,2017ApJ...843..129B,2017ApJ...835..113L,2018MNRAS.479.5184A,2019MNRAS.486.3805B}. Importantly, the morphologies and positions of the multiple images can also be used to constrain and uncover the underlying mass distribution of the lensing cluster, particularly in its most dense regions. Such mass reconstructions have been essential to our understanding of the distribution and properties of DM and have been shown to yield robust estimates of the masses in the cores of lensing clusters \citep[e.g.,][]{2005ApJ...621...53B,2007ApJ...668..643L,2013A&A...558A...1B,2014ApJ...797...48J,2014MNRAS.444..268R,2015ApJ...800...38G,2015MNRAS.452.1437J,2015ApJ...801...44Z,2016MNRAS.459.3447D,2016A&A...588A..99L,2017A&A...600A..90C,2019ApJ...873...96M}.  

Mapping the regions of greatest magnification is vital for the identification of cluster lenses that maximize the likelihood of discovering and characterizing lensed background sources. The lens-modeling techniques developed for this purpose can be broadly classified as parametric, non-parametric, or hybrids. In brief, parametric lens-modeling codes make use of physically motivated assumptions to model both galaxies and DM haloes analytically, whereas non-parametric modeling procedures reconstruct the underlying mass distribution directly on a grid of pixels. Requiring far fewer free parameters, the former approach tends to be more predictive than the highly flexible, non-parametric method (see \citealt{2013NewAR..57....1L} and \citealt{2017MNRAS.472.3177M} for a more thorough discussion and comparison of the different strategies). A common feature of all lens-modeling approaches is that the distances between the predicted and observed locations of lensed images are used to refine the models in a time-consuming, iterative process which requires substantial, prior knowledge of lensed images and their source redshifts. In addition, the resulting lens models are typically fine-tuned to optimize the model's fidelity to the specific cluster under investigation.

In this paper, we attempt to apply the same underlying principles to model cluster lenses ``blindly", i.e., without prior knowledge of multiple-image systems. Many parametric modeling techniques combine a smooth DM component, which describes the mass distribution on large scales, with small-scale mass perturbations at the locations of cluster galaxies, based on galaxy luminosities and assuming a constant mass-to-light ratio ($M/L$). \citet{2012MNRAS.423.2308Z}, however, showed that the size of the effective lensing area of massive clusters (i.e., the extent of the strong-lensing regime) can be estimated adequately by assuming that \textit{all} mass components are traced by the distribution of the light from cluster galaxies, a paradigm commonly referred to as ``light traces mass" (LTM). Using just ten calibration clusters previously analysed in depth using high-resolution \textit{Hubble Space Telescope} (\textsl{HST}) observations, Zitrin and co-workers developed and calibrated a straightforward algorithm to model 10,000 clusters found by ~\cite{2010ApJS..191..254H} in \textit{Sloan Digital Sky Survey} (SDSS) imaging. The resulting calibration is, however, highly specific to the characteristics of this data set, i.e., (1) the depth and image quality of the SDSS, (2) the constraints used in the study, and (3) conversions accounting for differences between \textsl{HST} and SDSS imaging. To model any other set of galaxy clusters using a similar method but different data requires a unique calibration and modifications specific to the dataset. Nonetheless, the simple prescription supplied by \citet{2012MNRAS.423.2308Z} for modeling galaxy clusters within the LTM paradigm has laid the foundation for cluster-lens detection and characterization from imaging data alone \citep{2018arXiv180703793C,2019MNRAS.482.1824S}.

We here employ the LTM principle in an effort to characterize the lensing strength of X-ray selected galaxy clusters discovered in the redshift range $z \sim 0.5$ to $z \sim 0.9$. To this end, we have developed a new LTM lens-modeling code, \texttt{AStroLens} (short for Automatic Strong-Lensing analysis), designed to map the strong-lensing regions of massive clusters with a common set of model parameters, and calibrated using unambiguously identified strong-lensing features in a few clusters from the same sample. In a companion paper (Ebeling et al., in preparation), we compare the performance of \texttt{AStroLens} using ground- and spacebased images of the same clusters, thereby exploring and quantifying systematic limitations and uncertainties introduced by the defining characteristics of different imaging data sets, such as field of view, depth, angular resolution, or photometric accuracy.

This paper is organized as follows. In Section \ref{sec:overview} we offer a brief overview of the principles of strong gravitational lensing for the non-expert. In Section \ref{eMACS} we introduce the sample of galaxy clusters used in this work, before describing all observations and the corresponding data reduction procedures in Sections \ref{obs} and  \ref{proc}, respectively. The subsequent photometric calibration of our data and the creation of the final source catalogues are the subjects of Section \ref{phot}. In Section \ref{id_gals} we explain the methods used to identify and select cluster-member galaxies. Section \ref{model_proc} outlines the LTM modeling procedure used by \texttt{AStroLens}, and in Section \ref{sec:par_cal} we define the minimization procedure followed to calibrate the free parameters of our model. We present the results of our minimization, predictions of the lensing strengths of our galaxy clusters, and comparisons between the lensing strength of eMACS clusters and other notable lenses from the literature in Section \ref{results}. The limitations of \texttt{AStroLens} and its uncertainties are discussed in Section \ref{sec:uncertainties}, and a summary of our main results is presented in Section \ref{sec:conclusion}. 

Throughout the paper we assume a flat $\Lambda$CDM cosmology with $\Omega_{m} = 0.3, \Omega_{\Lambda} = 0.7$, and $H_{0} = 100 h$ km s$^{-1}$ Mpc$^{-1}$, with $h = 0.7$.

%===================================================================================================
\section{Strong-Gravitational Lensing Overview} \label{sec:overview}
\begin{figure}
	\includegraphics[width=\columnwidth]{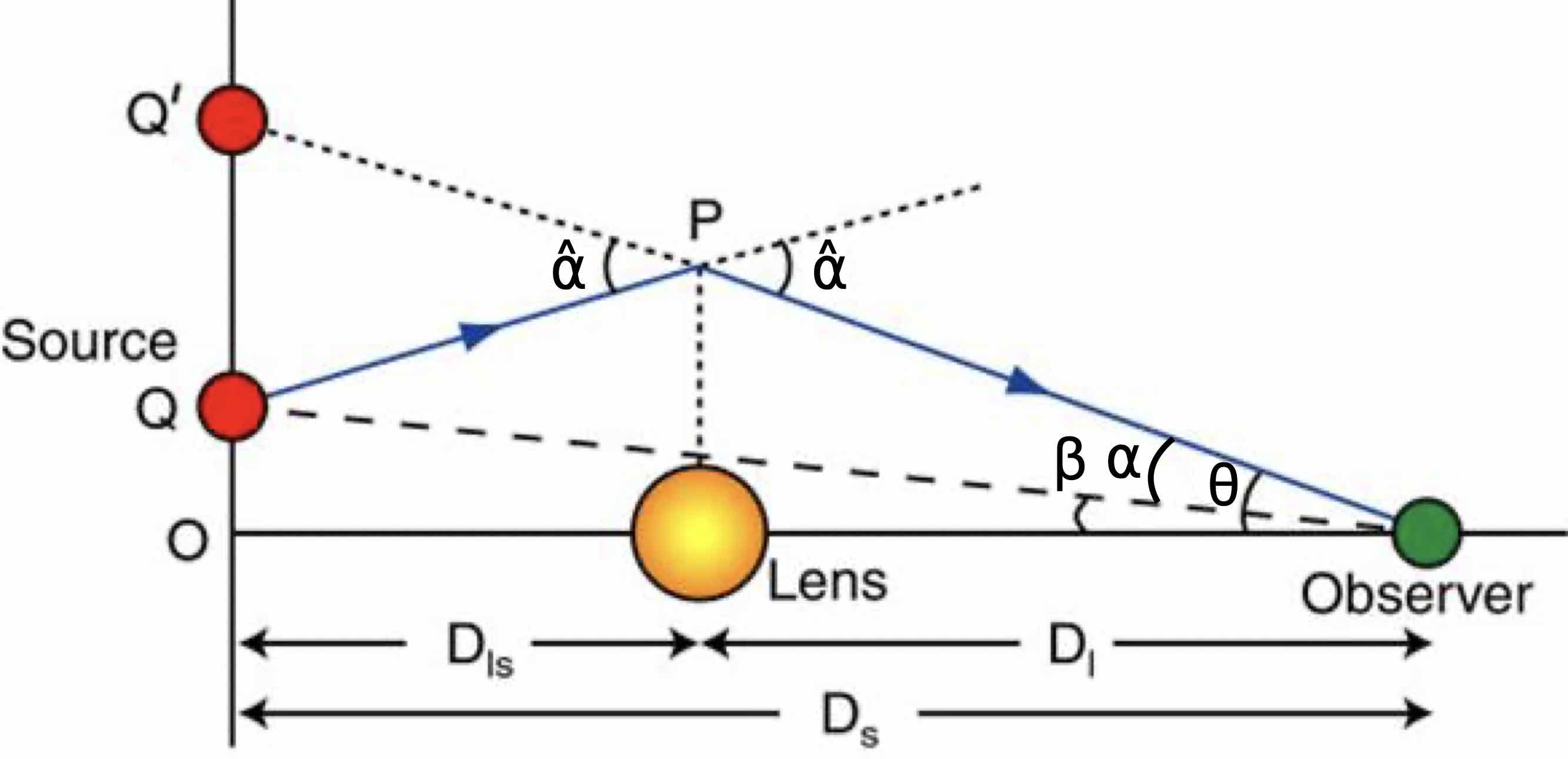}
    \caption{An idealized lensing system; angles are exaggerated for clarity. Light from the source is deflected by an angle $\alpha$, causing the  source's image to be observed at $\theta$ rather than $\beta$. (Figure credit: T. J. O'Brien \& N. J. Jackson, Jodrell Bank Observatory -- private communication)}
    \label{fig:diagram}
\end{figure}

The presence of any massive body (a ``lens") causes deflection of passing light rays. The amount of deflection (quantified by the deflection angle $\vec{\alpha}$) depends on the mass in the lens plane and the relative distances between the observer, the lens, and the light source. The lens equation ~\citep{1992grle.book.....S} parameterizes the deflection as the change of the object's position in the source and image planes:
\begin{equation} \label{eq:lens}
\vec{\beta}=\vec{\theta}-\vec{\alpha}(\vec{\theta}),
\end{equation}
where $\vec{\beta}$ is the true position of the source, $\vec{\theta}$ is the location where its image is observed, and $\vec{\alpha}(\vec{\theta})$ is the deflection angle. Fig.~\ref{fig:diagram} illustrates the path of a light ray in an idealized representation of a typical lensing system and defines the fundamental distances and angles used to quantify gravitational lensing.

\begin{figure}
	\includegraphics[width=\columnwidth]{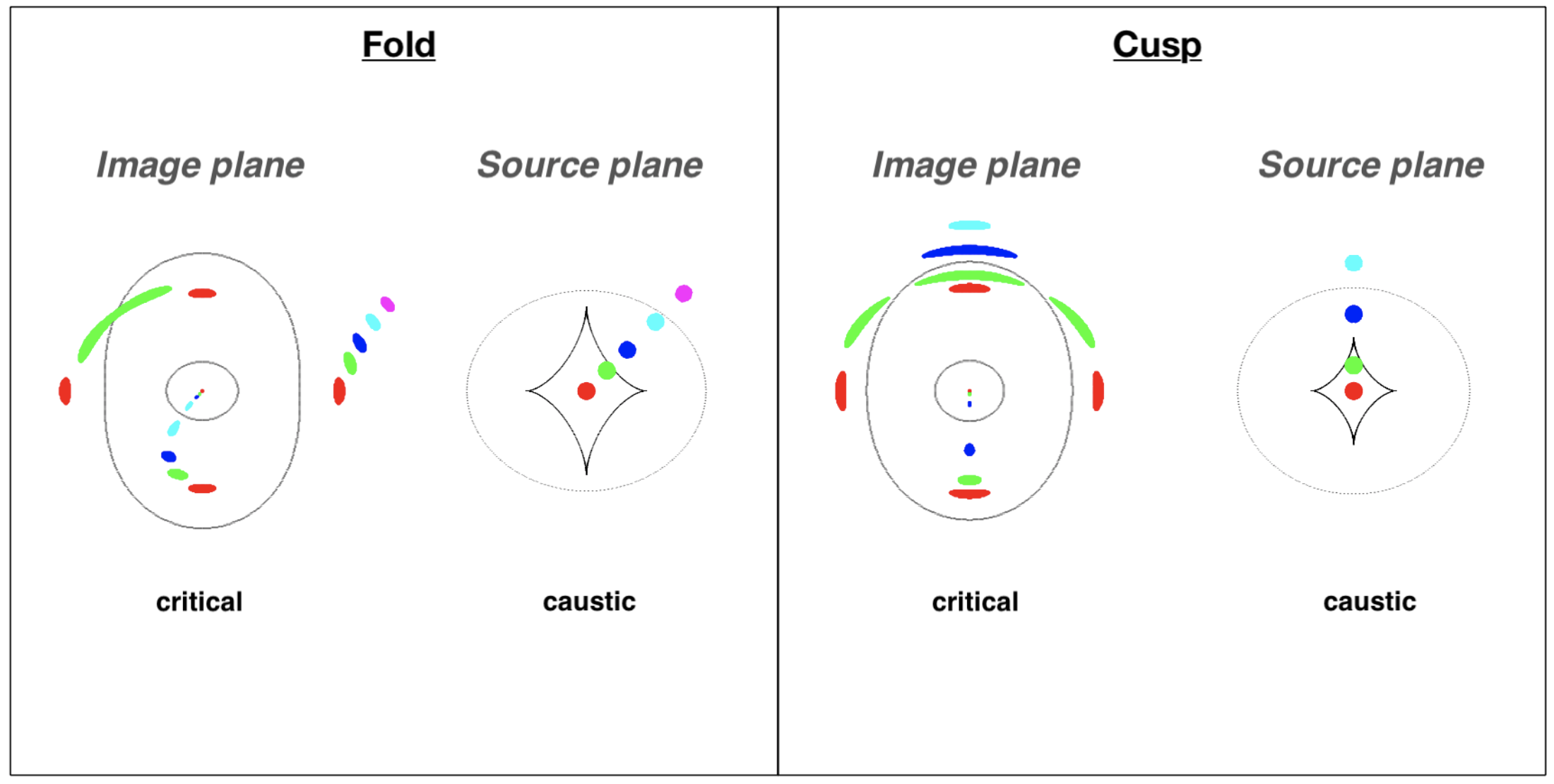}
    \caption{Positions and morphologies of lensed images with different alignments between a source and an idealized elliptical lens. The critical curves, which trace lines of infinite magnification around the area of greatest surface density, can be seen to pass through giant arcs and in between close pairs of multiple image systems. Caustics are the projections of critical lines onto the source plane. (Figure credit: \citealt{1996astro.ph..6001N})}
    \label{fig:diagram2}
\end{figure}

Strong-gravitational lensing can occur in all regions in the lens plane within which the surface-mass density exceeds a critical value given by
\begin{equation} \label{eq:crit}
    \Sigma\textsubscript{crit}=\frac{{\rm c}^{2}}{4 \pi {\rm G}}  \frac{d\textsubscript{s}}{d\textsubscript{l} d\textsubscript{ls}}.
\end{equation}
Here G is the universal gravitational constant, c is the speed of light, and $d\textsubscript{l}, d\textsubscript{s},$ and $d\textsubscript{ls}$ are the angular-diameter distances from the observer to the lens, from the observer to the source, and from the lens to the source, respectively. For an extended source, the deflection angle can be found analytically by convolving the convergence $\kappa \equiv \frac{\Sigma}{\Sigma_{\rm{crit}}}$ (i.e., the dimensionless surface mass density) with the kernel function $\frac{1}{\pi} \frac{\theta}{|\theta|}$:
\begin{equation}
    \alpha(\theta) = \frac{1}{\pi} \int \kappa(\theta^\prime)  \frac{\theta-\theta^\prime}{|\theta-\theta^\prime|} d^2 \theta^\prime,
\end{equation}
where $\theta$ is the angular distance from a point in the image plane to a given surface mass element ~\citep{2011A&ARv..19...47K}.  Equivalently, the projected surface mass density can be derived from the divergence of the deflection field through
\begin{equation} \label{div_alpha}
    \vec{\nabla} \cdot \vec{\alpha} = 2 \kappa.
\end{equation} 
In general, when $\kappa > 1$, the deflection field yields multiple paths between a source and its image in the plane of the lens, thereby causing multiple images of a single source to be observed.

The lens equation (Eq.~\ref{eq:lens}) quantifies how images are mapped from the source plane to the lens plane, and it is this mapping that creates the distortions and magnifications that characterize the strong-lensing regime. A schematic overview of typical strong-lensing configurations is shown in Fig.~\ref{fig:diagram2} for an assortment of alignments between a source and an idealized elliptical mass distribution. Critical lines (virtual lines of infinite magnification) form in regions within which the surface density exceeds the critical surface density defined in Eq.~\ref{eq:crit}. Any image mapped to a location close to a critical line is greatly magnified and part of a multiple-image system which, depending on the alignment between the source and the lens, can be merged into a giant arc or observed as separate images. For the purposes of this work, it is important to note that giant arcs trace the critical line to good approximation (see the "cusp" arc configuration in Fig.~\ref{fig:diagram2}), and that multiple-image systems frequently include close pairs that straddle the critical line. Recognizing these characteristic strong-lensing features is critically important for the identification of constraints for lens modeling. 

Another powerful global property widely used to characterize strong-lensing systems is the effective Einstein radius, defined as the equivalent radius of the area $A$ (also known as the critical area) enclosed by the critical lines, i.e., $\theta\textsubscript{E} = \sqrt{A/\pi}$. For spherically symmetric lenses, the average surface-mass density of the critical area is equal to $\Sigma\textsubscript{crit}$, the value of which depends on the redshifts of the lens and the source. By measuring $\theta\textsubscript{E}$ one can infer the power of the lens, i.e., the size of the critical area, and thus the probability of detecting highly magnified images of background sources. In addition, $\theta\textsubscript{E}$ directly yields an estimate of the mass within the critical area, as, for a point-like mass and a source that are perfectly aligned along our line of sight, the Einstein radius 
is given by
\begin{equation}
    \theta\textsubscript{E}=\sqrt{\frac{4{\rm G}M(< \theta\textsubscript{E})}{{\rm c}^2}\frac{d_{\rm{ls}}}{d_{\rm{s}} d_{\rm{l}}}}.
	\label{eq:Einstein Radius}
\end{equation}
Lastly, the distribution of Einstein radii across the sky and their relative sizes can be used to understand the geometry of the Universe and 
even as a probe of the primordial non-Gaussianity of the cosmic density field 
~\citep{2009MNRAS.392..930O}. 

%===================================================================================================
\section{Cluster sample} \label{eMACS}
Building on the success of the MAssive Cluster Survey \citep[MACS,][]{2001ApJ...553..668E}, which increased the number of known extremely X-ray luminous galaxy clusters beyond $z = 0.3$ by a factor of 30, the extended Massive Cluster Survey (eMACS) used X-ray detection coupled with optical confirmation to discover massive galaxy clusters at even higher redshifts of $z \geq 0.5$ ~\citep{2013MNRAS.432...62E}. Covering over 21,000 square degrees, eMACS has, to date, compiled a sample of 99 clusters out to redshift $z = 0.9$, not counting 12 clusters at $z>0.5$ identified before by the MACS project \citep{2007ApJ...661L..33E} and re-discovered by eMACS. All eMACS clusters were confirmed by screening optical images obtained during the Pan-STARRS1 ~\citep{2016arXiv161205560C} $3\pi$ Survey for galaxy overdensities at the locations of all spectrally hard X-ray sources detected in the ROSAT All-Sky Survey (RASS; ~\citealt{1999A&A...349..389V}) within the survey footprint. For further details of the eMACS survey design and galaxy cluster detection methods, see \cite{2013MNRAS.432...62E} and Ebeling et al.~(in preparation).

The sample used in this work consists of 96 eMACS clusters for which ground-based imaging was obtained by us from Maunakea (HI). All spectroscopic information on our target clusters (including measurements of redshifts and velocity dispersions) as well as X-ray luminosities were taken from the eMACS database (see Ebeling et al., in preparation). 

%===================================================================================================
\section{Observational data} \label{obs}
Our analysis is based on shallow imaging obtained with the \textit{Gemini} Multi-Object Spectrograph on the 
\textit{Gemini-North Observatory} (GMOS-N; ~\citealt{2004PASP..116..425H}). In addition, we used archival, high-resolution \textsl{HST} images of eMACS clusters, obtained with the Advanced Camera for Surveys \citep[ACS,][]{1998SPIE.3356..234F} and the Wide-Field Camera 3 (WFC3) for the SNAPshot programs GO-13671, -14098, -15132, and -15843 (PI Ebeling) to aid in the identification of strong-lensing features used to calibrate our LTM algorithm. We use the same \textsl{HST} images in Ebeling et al.\ (in preparation) to explore and quantify  systematic effects affecting LTM analyses of galaxy clusters as well as uncertainties specific to ground- or space-based imaging data.

\subsection{Gemini-North GMOS Imaging}
Ground-based images of 96 eMACS clusters were obtained with GMOS-N between 2015 and 2018 in the $g', r',$  and $i'$ filters (one cluster, eMACSJ0324.0, was observed only in $r'$ and $i'$). %The GMOS filters were designed to yield transmission functions nearly identical to those used by the SDSS. 
In each filter, three to four dithered 90-second exposures were obtained. With a field of view covering 5.5 square arcminutes (roughly 4 Mpc$^2$ for a cluster at $z = 0.6$), the GMOS-N images capture the full extent of the mass distribution in all but the most extended merging clusters. %Unfortunately, these final image products are at times noisy. 
%We reduce all GMOS-N imaging data with the Gemini IRAF package, following standard procedures as described in Section \ref{gmosProc}. 
Table ~\ref{tab:log_obs} lists the program I.D.s and the detectors used in these observations.

\subsection{Keck spectroscopy}
\label{sec:spec}

Spectroscopic follow up of eMACS clusters was conducted at optical and near-infrared wavelengths using the MOSFIRE and DEIMOS spectrographs on the Keck-1 and -2 10m telescope, respectively, and targeted both presumed cluster members and strong-lensing features. Observational details can be found in \citet{2013MNRAS.432...62E} and Ebeling et al.\ (in preparation). The redshifts collected in this multi-year observing campaign are used here to improve the selection of cluster members (see Section~\ref{sec:rs_select}) and to compile a set of confirmed strong-lensing features in eMACS clusters, specifically multiple-image sets and giant arcs, for the critical calibration of \texttt{AStroLens}' global parameters, described in Section~\ref{sec:par_cal}.

\begin{table}
    \centering
    %\topcaption{Log of Observations. Note that the sum of clusters presented here exceeds 94, as some clusters were observed on multiple occasions and 
    %two eMACS sources were later found to not be galaxy clusters.} 
    \caption{Log of \textit{Gemini}/GMOS observations. Note that the total number of clusters observed exceeds 96, as some targets were observed multiple times.} %Conversely, two clusters were eliminated after follow-up study showed the RASS X-ray emission to originate primarily from point sources.} 
    \begin{tabular}{llc} % Column formatting, @{} suppresses leading/trailing space
       \multicolumn{3}{l}{} \\
       \hline 
       \hline
        No. clusters & GMOS-N Observation I.D. & CCD Array  \\     
       \hline
       3 & GN-2015A-FT-20 & e2v \\   
       17 & GN-2015A-Q-25 & e2v \\
       4 & GN-2015B-FT-5 & e2v \\
       4 & GN-2015B-FT-13 & e2v \\
       25 & GN-2015B-Q-42 & e2v \\
       19 & GN-2016A-Q-60 & e2v \\
       12 & GN-2016B-Q-8 & e2v \\
       1 & GN-2017A-Q-21 & e2v \\
       17 & GN-2018A-FT-101 & Hamamatsu \\
       8 & GN-2018B-Q-112 & Hamamatsu  \\
       \hline
    \end{tabular}
    \label{tab:log_obs}
 \end{table}
%===================================================================================================
\section{Data Processing} \label{proc}
We follow standard reduction procedures to produce science-quality images of the \textit{Gemini} observations, as summarized below. 

Raw images are first bias-subtracted using a master bias frame created by combining at least 50 bias frames taken within 1-2 months of the target observation. We also create master flat-field frames from at least 20 individual twilight frames in each band pass to correct for pixel-to-pixel variations in the detector sensitivity. In addition, we include corrections calculated from a standard overscan region in the detector. Flat-fielding and bias subtraction are carried out using the IRAF routine \texttt{gireduce}. We note that a fringe correction was required in each bandpass (not just the $i'$-band) to obtain distortion-free final images (credit: Gemini Help Desk). This correction is calculated with the IRAF routine \texttt{gifringe} and applied with the routine \texttt{girmfringe}. Without this correction, the final images are plagued by artifacts and excess flux around bright sources. The final fringe-corrected frames from the GMOS-N detector's three individual CCDs are combined into a single mosaic using the Gemini IRAF routine \texttt{gmosaic}, sky-subtracted, and cleaned of cosmic rays before being co-added with the IRAF routine \texttt{imcoadd}. In a final step all images for a given target are registered to each other and a common world-coordinate system.

\begin{figure*}
	\includegraphics[width=\textwidth]{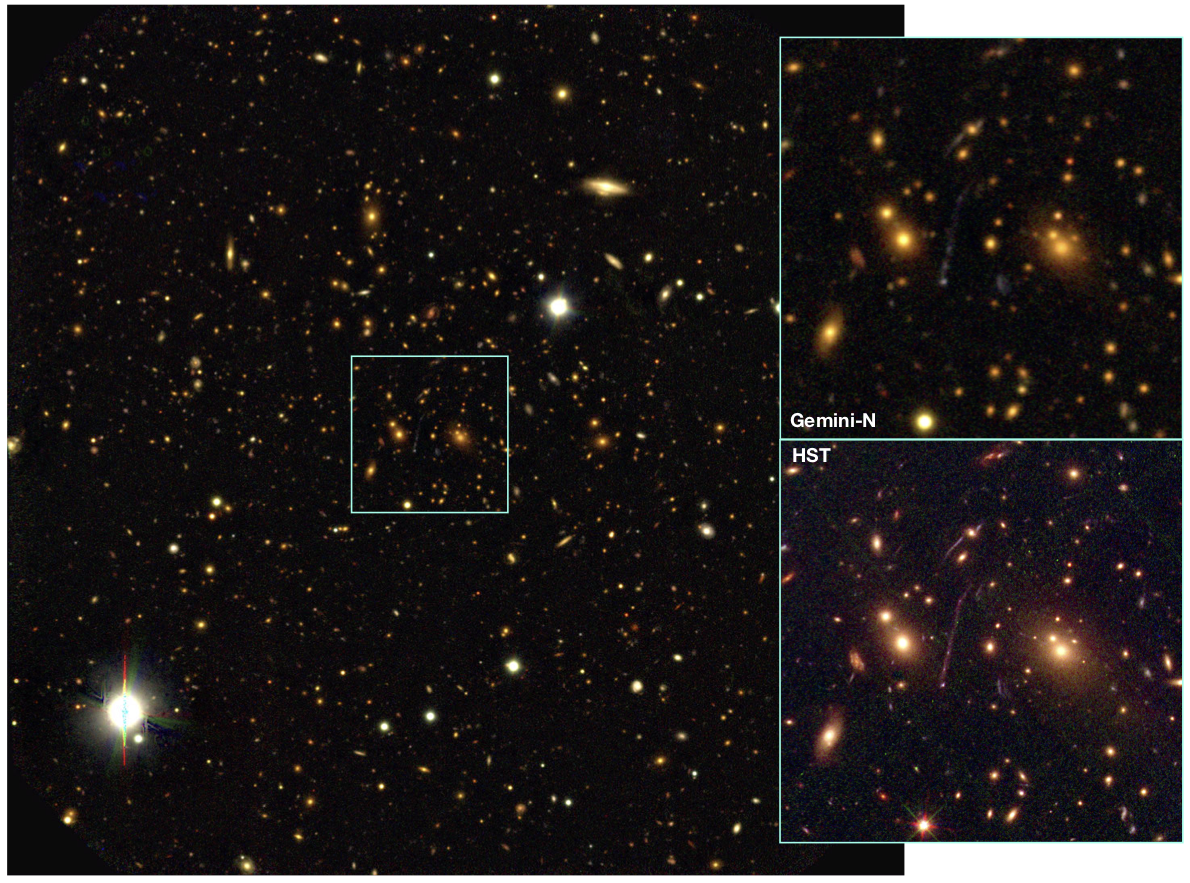}
    \caption{eMACSJ2316.6, a massive galaxy cluster at $z=0.526$, as viewed with \textit{Gemini}/GMOS (filters: g$^\prime$, r$^\prime$, i$^\prime$). The highlighted area centred on the cluster core (1 arcmin on a side; 376 kpc at the cluster redshift) contains several strong-lensing features. The zoomed-in images of the same central area, shown in the insets, illustrate the advantages of space-based imaging for the identification of multiple-image systems (\textit{HST} filters: F606W, F814W, F110W).
    }\label{fig:EMACS2316.6}
\end{figure*}

For each filter, the final image is the average of a stack of three to four 90-second exposures and is binned $2\times2$. The final GMOS-N images have a
pixel scale $\sim 0.14" /$pixel. An example is shown in Fig. \ref{fig:EMACS2316.6} (top).

%===================================================================================================
\section{Source Catalogues} \label{phot}
To generate source catalogues, we use SExtractor version 2.19.5 ~\citep{1996A&AS..117..393B} in dual-image mode, adopting the reddest passband, i$^\prime$, as the detection band.  In addition to source positions and photometric measurements in each passband (see Section~\ref{sec:phot}), we also record each source's ellipticity, orientation angle, and peak surface brightness in the detection band. Corrections for Galactic extinction are assigned to each source based on the dust maps provided by ~\cite{2011ApJ...737..103S}. We use the default SExtractor settings except for a detection threshold of $5\sigma$ in surface brightness, a minimal detection area of 16 pixels, and a minimal deblending contrast of 0.0001. The cited values mitigate blending of individual sources in crowded areas without breaking up extended sources. 

\subsection{Photometry}
\label{sec:phot}

We record SExtractor's \texttt{mag\_auto} values to measure total magnitudes in elliptical apertures for extended sources (and circular apertures otherwise) out to 2.5 times the Kron radius \citep{1980ApJS...43..305K}.

Photometric zero points are derived manually by matching our source catalogues to stars in the Pan-STARRS1 DR2 database\footnote{We note that colour transformations between the PS1 filters and the SDSS filters result in differences smaller than our instrumental error ~\citep{2012ApJ...750...99T}, and that the colour transformations between the SDSS and GMOS-N filters are similarly modest \citep{2009PASA...26...17J}. While such colour transformations could have slight effects on the selection of cluster galaxies (Section ~\ref{sec:rs_select}), the results presented here are otherwise insensitive to them.} \citep{2016arXiv161205242M}. If all \textit{Gemini} observations had been taken in photometric conditions, the zero points thus derived would be the same for all fields, except for minor variations caused by instrumental effects, such as dust buildup or slow degradation of the detector sensitivity. In practice, however, residual moon light, thin clouds, and other observational characteristics and conditions introduce small differences between the photometric zero points for different observations. 

A comparison of the zero points derived from PS1 and SDSS photometry for the 60 clusters in our sample that fall inside the SDSS footprint shows that only very few cluster fields were observed in non-photometric conditions, leading to abnormally high zero points; only r$^\prime$-band observations are affected. The scatter in the remaining zero points of about 0.06 mag ($1\sigma$) is the primary source of uncertainty both for the colour-based selection of likely cluster members (see Section~\ref{sec:rs_select}) and the computation of galaxy luminosities. The comparison between PS1 and SDSS photometry supports this assessment, as it shows a similar scatter of 0.07 mag for either passband, along with a slightly smaller systematic difference of about 0.05 mag between PS1 and SDSS.

\subsection{K corrections}
\label{sec:kcorr}

\begin{figure}
	\includegraphics[width=\columnwidth]{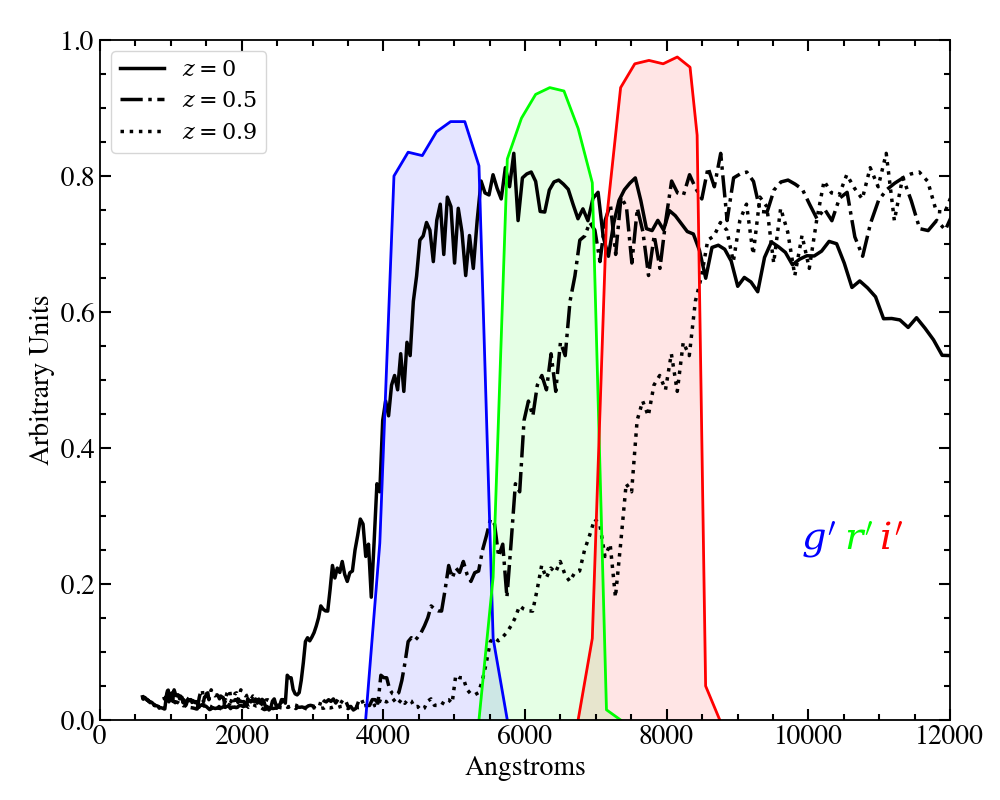}
    \caption{Early-type galaxy spectrum at $z = 0, z = 0.5,$ and $z = 0.9$ superimposed on the transmission functions of the broad-band filters used in this work. Note that the 4000\AA\ break falls into the \textit{Gemini} r$^\prime$ and i$^\prime$ bands for the redshift range covered by our cluster sample.}
    \label{fig:kcorrect}
\end{figure}

Since, in this application, our LTM algorithm uses the light from galaxies in clusters across a range of redshifts as its primary input, observed apparent magnitudes must first be converted to intrinsic luminosities, thereby accounting for differences in redshift between our cluster lenses. This conversion involves not merely a shift along the magnitude axis (effectively the subtraction of a suitably chosen distance modulus), but also a colour correction (``k correction") that compensates for the redshifting of light out of the chosen observational passband. Fig.~\ref{fig:kcorrect} illustrates this effect by showing the spectrum of an early-type galaxy at increasing redshifts, along with the transmission functions of the \textit{Gemini} filters used by us.

We compute k corrections for each cluster using the \texttt{kcorrect} package developed by \citet{2007AJ....133..734B} to adjust observed magnitudes from the actual cluster redshift to a universal fiducial redshift. For the selection of likely cluster members (see Section ~\ref{sec:rs_select}), we choose a reference redshift of $z = 0.6$, while for absolute magnitudes and luminosities we apply k corrections to $z = 0$. Since the spectral energy distributions (SED) of individual galaxies in our fields are unknown, we adopt the template of a canonical early-type SED (shown in Fig.~\ref{fig:kcorrect}) for all galaxies.

%===================================================================================================
\section{Identification of Cluster Galaxies} \label{id_gals}

\begin{figure}
	\includegraphics[width=\columnwidth]{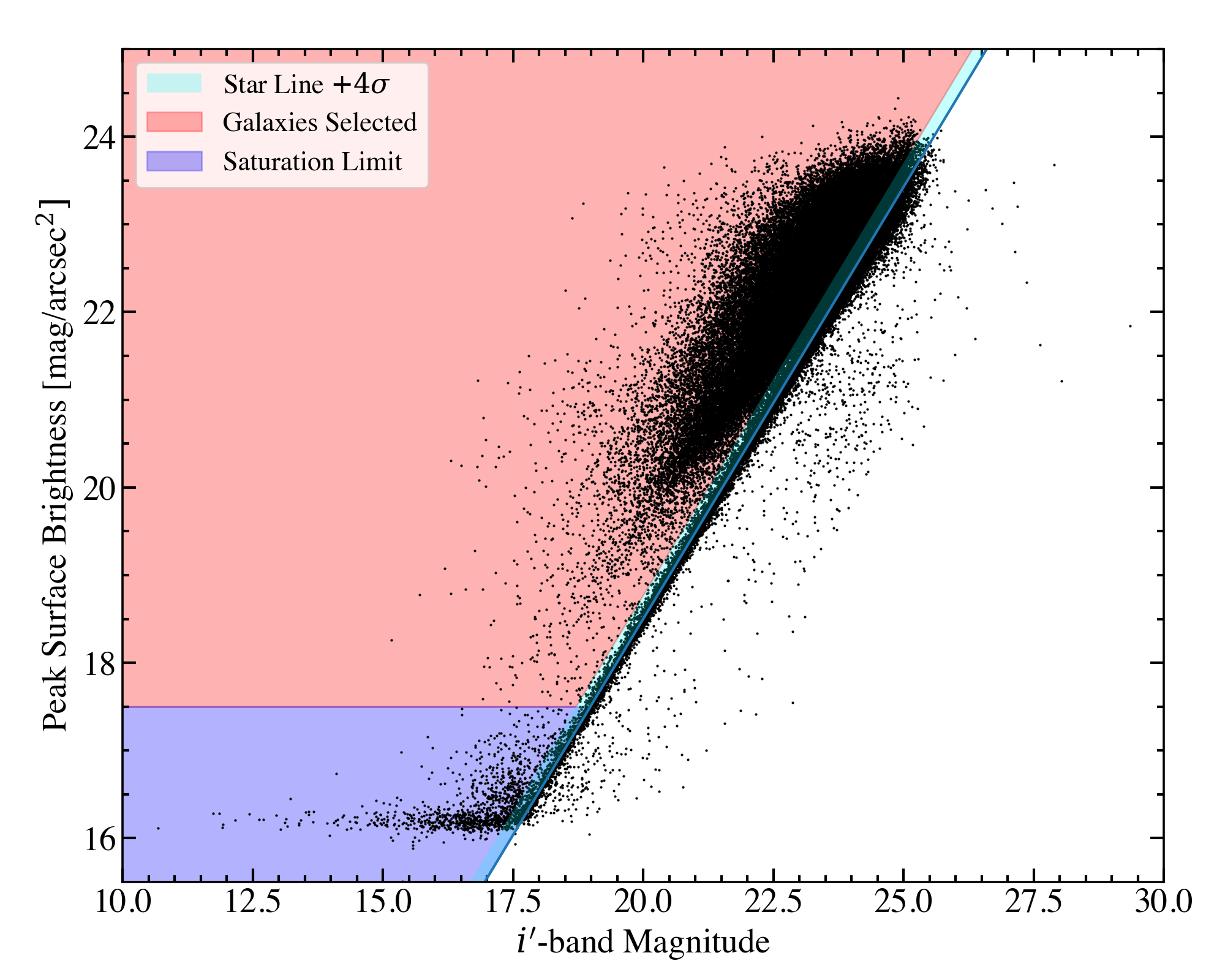}
    \caption{Peak surface-brightness versus magnitude for all sources detected in our \textit{Gemini} observations. Stars follow a linear relationship until their flux reaches the detector's saturation limit. Sources below the star line are excluded as artifacts, whereas sources falling above the $+4\sigma$ limit of the star line are designated galaxies.}
    \label{fig:stargal}
\end{figure}

In order to map and measure the distribution of cluster light, we need to  be able to distinguish cluster member galaxies from contaminating stellar sources, as well as from background and foreground galaxies.

\subsection{Star-Galaxy Separation}
Although SExtractor computes a \texttt{class\_star} value, meant to quantify the probability of a source being stellar, we found this parameter to be unreliable for our shallow \textit{Gemini} images. We therefore resort to using our own star-galaxy separation procedure which exploits the fact that all stellar sources are scaled images of the same instrumental point-spread function; as such, their peak surface brightness scales linearly with their total flux. By contrast, the diverse morphology of galaxies, expressed in a wide range of surface-brightness profiles, ellipticities, and substructure, causes galaxies to deviate from this linear scaling relation. Fig.~\ref{fig:stargal} demonstrates this difference between stellar sources and galaxies when all sources are stacked\footnote{The stacking is performed after shifting the data for each field to an arbitrary common reference point in i$^\prime$ magnitude to account for differences in spatial resolution, i.e., seeing.}. 

We fit the relation for stellar sources with a line of slope unity and flag as stars all sources within a band around the star line or below the saturation limit shown in Fig.~\ref{fig:stargal}. Sources below the star line are unphysically compact and classified as artifacts. The remaining sources, i.e., all unsaturated sources lying at least $4\sigma$ above the best-fit star line, are considered to be galaxies.

\subsection{Red-Sequence Selection} \label{sec:rs_select}
Given the small number of broad-band filters used in this study, cluster membership can not credibly be established from photometric redshifts. We can, however, exploit a physical characteristic of galaxies in high-density environments, the cluster ``red sequence" ~\citep{1977ApJ...216..214V}, a prominent feature in the colour-magnitude diagram (CMD) of clusters that is created by the fact that early-type galaxies, which represent the vast majority of cluster galaxies, have ceased to form stars, resulting in an almost uniform colour that varies only slightly with galaxy luminosity. By contrast, fore- and background sources, being at a different redshift and having different intrinsic colours than cluster ellipticals, do not follow the red sequence. The red-sequence selection method is well tested and has been applied successfully to select elliptical galaxies within $\Delta z \sim 10$\% ~\citep{2000AJ....120.2148G}, provided the underlying photometry is of high quality. 

Since the very existence of the red sequence in a CMD is predicated on the chosen colour capturing the only prominent spectral feature of elliptical galaxies (the 4000\AA\ break in the continuum), no set of two filters can reliably bring out the red sequence in clusters at all redshifts. With only three passbands used in our \textit{Gemini} observations (g$^\prime$, r$^\prime$, i$^\prime$), the two filters that approximately bracket the 4000\AA\ break across most of our sample's redshifts range ($z=0.5-0.9$) are r$^\prime$ and i$^\prime$ (Fig.~\ref{fig:kcorrect}), making r$^\prime$--i$^\prime$ the colour of choice for the construction of suitable CMDs. 

To select likely cluster members in a uniform manner for all clusters in our sample, we subtract the relative distance modulus between the cluster redshift and $z=0.6$ from the observed i-band magnitudes of all galaxies, thereby effectively moving all clusters to the fiducial redshift of $z=0.6$; in addition, we apply k-corrections from the observed redshifts to $z = 0.6$. By doing so, we account for differences in observed magnitudes caused by the redshifting of light out of a given photometric bandpass (see Section~\ref{sec:kcorr}) and allow uniform sampling of each cluster's intrinsic luminosity function. The resulting standardized colour-magnitude diagrams still show slight variations in depth, as well as in the location of the red sequence, which are caused by a combination of (a) only marginally photometric conditions during the respective observations, (b) photometric uncertainties introduced by slight differences between the GMOS and PS1 filter sets, and (c) physical, redshift-dependent evolution of the cluster red sequence. We account for all three effects by shifting each cluster's CMD in colour space such that all red sequences align. The required shifts are small; the average shift in r$^\prime$--i$^\prime$ is $-0.01$, with a $1\sigma$ standard deviation of 0.09.

\begin{figure}
	\includegraphics[width=\columnwidth]{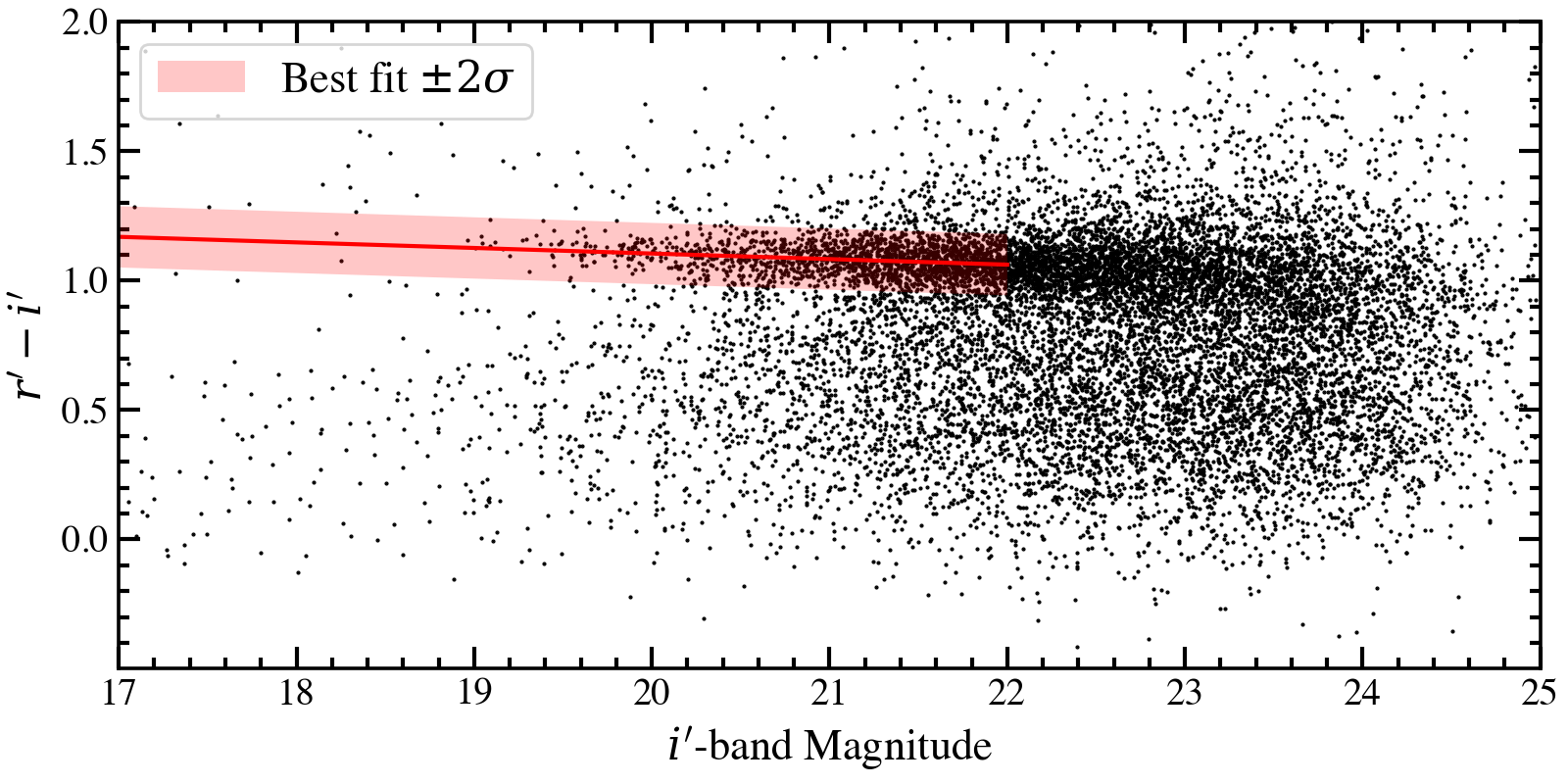}
    \caption{Colour-magnitude diagrams for all galaxies after adjustment of all magnitudes to a fiducial cluster redshift of $z=0.6$. We fit a straight line (with a Gaussian density profile) to the red sequence and select all galaxies within $\pm 2\sigma$ (and with $m_{i^\prime}<22$) as likely cluster members.}
    \label{fig:RS_S}
\end{figure}

The resulting, stacked CMD is shown in Fig.~\ref{fig:RS_S}, together with the best-fitting linear description of the red sequence and its $2\sigma$ width. We assume that all galaxies within the shown red-sequence band are cluster members, but add to this set all galaxies outside this colour range that have been spectroscopically confirmed as cluster members in the follow-up observations described in Section~\ref{sec:spec}; conversely, we remove all galaxies spectroscopically identified as fore- or background sources.

%===================================================================================================
\section{AStroLens} \label{model_proc}

Our algorithm to model cluster lenses builds on the successes of many previous studies \citep[e.g.,][]{2005ApJ...621...53B, 2009MNRAS.396.1985Z, 2011MNRAS.417..333M, 2011MNRAS.410.1939Z, 
2011MNRAS.413.1753Z, 2011ApJ...742..117Z, 2015ApJ...801...44Z, 2019MNRAS.482.1824S, 2019arXiv190509802C} that make use of the LTM assumption that the mass distribution within a galaxy cluster, from both the galaxies and the cluster-scale DM halo, is approximately traced by the distribution of light. Indeed, luminous red galaxies have been shown to be powerful tracers of the mass concentrations both individually \citep{2009ApJ...707..554Z} and in superposition along the line of sight \citep{2013ApJ...769...52W}. We here model the deflection of light caused by two mass components: individual, relatively compact galaxies and, on much larger scales, extended DM halos. 

Although it follows the same fundamental  paradigm, our approach differs from previous LTM algorithms \citep[e.g.,][]{2012MNRAS.423.2308Z,2019MNRAS.482.1824S} in several important ways that affect the modeling of both the galaxy- and cluster-scale components, as described in detail in the following sections. Specifically, \texttt{AStroLens} differs from previous LTM algorithms in the choice of mass profile used to describe both cluster galaxies and the cluster-scale DM halo, and also uses a unique approach to determine the appropriate geometry and structure of the cluster-scale DM halo for each cluster.

Using only three free parameters whose values are calibrated with the help of known cluster lenses with spectroscopically confirmed sets of multiple images and/or giant arcs (see Section~\ref{sec:par_cal}), \texttt{AStroLens}  predicts the locations of critical lines in an automated fashion from solely the positions, luminosities, and geometries of the galaxies in a cluster lens. From the critical lines, we estimate each cluster's Einstein radius and mass within the critical area. We note that the values of the aforementioned model parameters can be straightforwardly calibrated for any cluster sample and imaging data set that provides sufficient strong-lensing constraints.

\begin{figure*} %\h
	\includegraphics[width=\textwidth]{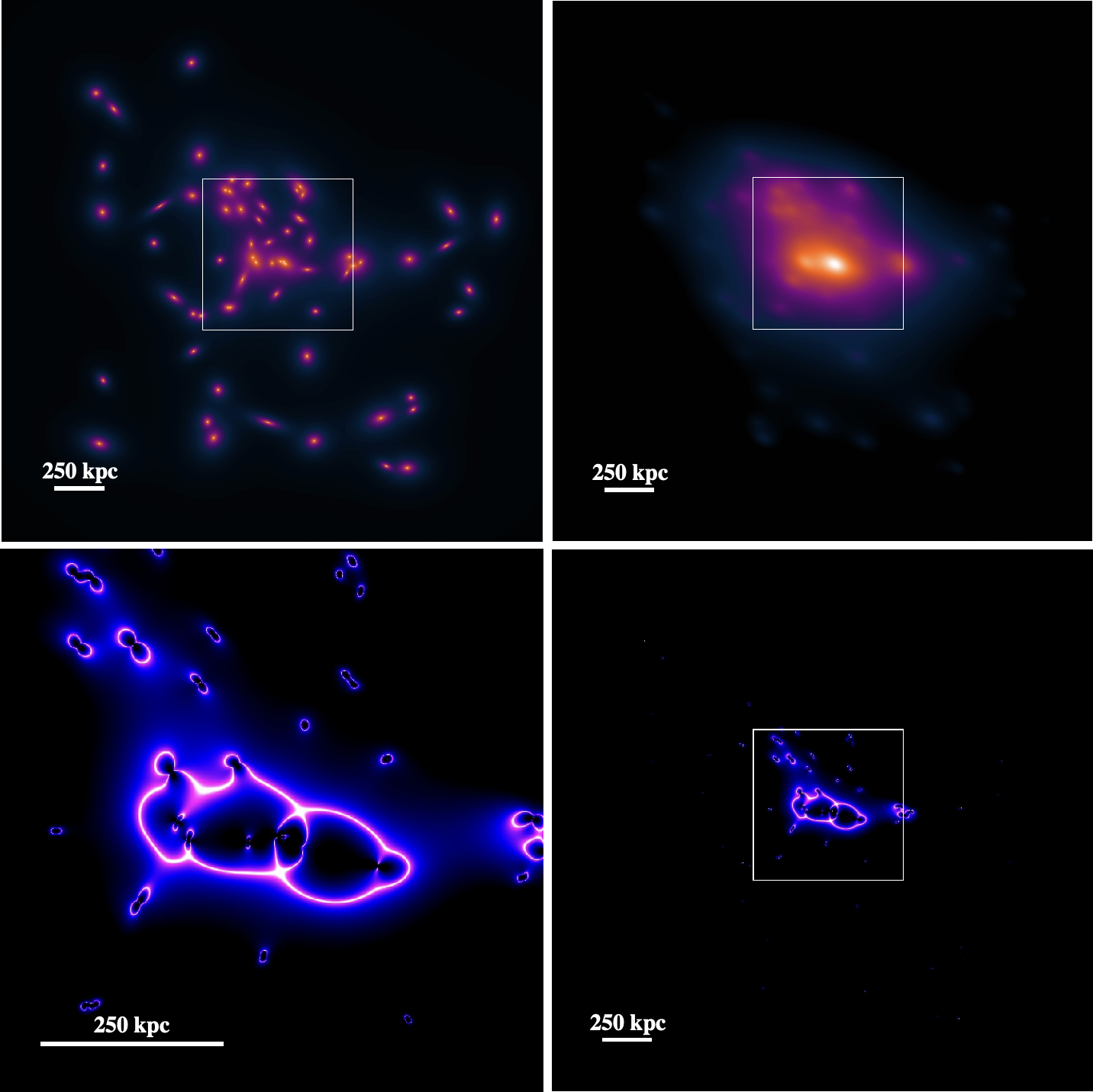}
    \caption{Key data products illustrating the \texttt{AStroLens} processing of the \textit{Gemini} data for eMACSJ2316. \textit{Clockwise from top left}: Galaxy-scale component of the mass distribution; cluster-scale component of the mass distribution, obtained by convolving the galaxy light distribution with a PIEMD kernel of size $R\textsubscript{core}$ (see Section~\ref{sec:clus_scale} for details); magnification field; and close-up view of the magnification field within the box marked in the other panels of this figure (the bright pink lines mark the locations of infinite magnification).}
    \label{fig:gal}
\end{figure*}

\subsection{Galaxy-Scale Component} \label{sec:gals}

Our modeling of the galaxy-scale component of the mass distribution follows the basic principles outlined in ~\cite{2012MNRAS.423.2308Z} but differs in several key aspects. Whereas in previous LTM work all cluster galaxies are modeled with a spherically symmetric radial density profile that follows a universal power law, we assign each galaxy a pseudo-isothermal elliptical mass distribution (PIEMD), whose surface-density profile is given by
\begin{equation}\label{eq:piemd}
    \Sigma(r)=\frac{S_{0} r_{\rm{cut}}}{2 \mathrm{G}\left(r_{\rm{cut}}-r_{\rm{core}}\right)}\left(\frac{1}{\sqrt{r_{\rm{core}}^{2}+r^{2}}}-\frac{1}{\sqrt{r_{\rm{cut}}^{2}+r^{2}}}\right).
\end{equation}
This is the same profile as used by most state-of-the-art parametric lens modeling packages \citep[e.g.,][]{2005MNRAS.356..309L,2007NJPh....9..447J}. In Eq.~\ref{eq:piemd} $S_{0}$ is a normalization factor (with units of velocity dispersion squared), and the boundaries $r\textsubscript{core}< r < r\textsubscript{cut}$ define the radial range over which the mass distribution is isothermal
~\citep{2007NJPh....9..447J}. Accordingly, the PIEMD profile has seven parameters: in addition to $r\textsubscript{core}$, $r\textsubscript{cut}$, and $S_{0}$ defined above, $x$ and $y$ specify the position of the galaxy, and $e$ and $\theta$ describe its ellipticity and orientation, respectively. The latter four quantities are directly measured by SExtractor (Section ~\ref{phot}). This leaves three free parameters, $S_{0}$, $r_{\rm{core}}$, and $r\textsubscript{cut}$.

The PIEMD profile can be integrated to obtain the mass interior to $\theta$, 
\begin{equation}\label{eq:piemdmass}
%    M(<\theta)=\frac{L}{L_\odot}\frac{2 \pi K}{(2-q)}\left(d_{\rm{l}} \theta\right)^{2-q},
    M(<\theta)=\frac{\pi S_0 r_{\rm{cut}}}{\mathrm{G}}\left( 1- \frac{\sqrt{r_{\rm{cut}}^2+(d_{\rm{l}} \theta)^2} -\sqrt{r_{\rm{core}}^2+(d_{\rm{l}} \theta)^2}}{r_{\rm{cut}}-r_{\rm{core}}}\right)
\end{equation}
where $\theta$ is the angular distance from the centre of the galaxy, and $d_{\rm{l}}$ is the angular-diameter distance from the observer to the lensing cluster. The deflection field created by a single galaxy is given by
\begin{equation} \label{eq:alpha}
\alpha(\theta)=\frac{4 G M(<\theta)}{c^{2} \theta} \frac{d_{l s}}{d_{\rm{s}} d_{\rm{l}}}.
\end{equation}
Using Eq.~\ref{eq:piemdmass}, Eq.~\ref{eq:alpha} becomes
\begin{equation} \label{alpha_expand}
%\alpha_{\rm gal}(\theta)=\frac{8 \pi G}{c^2} \frac{L}{L_\odot} \frac{K}{2-q} \frac{d_{l s}}{d_{\rm{s}}} (d_{\rm{l}}\theta)^{1-q}.
    \alpha_{\rm gal}(\theta) = \frac{4\pi S_0 r_{\rm cut}}{c^2 \theta}
    \left( 1- \frac{{\sqrt{r_{\rm{cut}}^{2}+(d_l\theta)^{2}}}-{\sqrt{r_{\rm{core}}^{2}+(d_l\theta)^{2}}}}{r_{\rm cut}-r_{\rm{core}}}\right)
     \frac{d_{ls}}{d_s d_l}
\end{equation}
(We note at this point that we compute $\alpha_{\rm gal}$ not via the above equation but by convolving $\Sigma(r)/\Sigma_{\rm crit}$ with the kernel function $\smash{\frac{1}{\pi}\frac{\theta}{|\theta|}}$, as explained in Section~\ref{sec:overview}.) 

So far, we have only quantified the amplitude of the deflection. However, as shown in Eq.~\ref{eq:lens} and in Fig.~\ref{fig:diagram}, the deflection is in fact a vector in the source-lens-observer plane. Hence, the deflection field created by all cluster galaxies is the vector sum of their individual deflection fields. If $\vec{\theta}$ represents a particular point in two-dimensional space, the full deflection field induced at that point by cluster galaxies located at $\vec{\theta}_{i}$ is given by
\begin{equation} \label{eq:1}
\vec{\alpha}\textsubscript{gal}(\vec{\theta}) = \sum_{i} \alpha_{\rm gal,i}(\theta) \frac{\vec{\theta}-\vec{\theta}_{i}}{\left|\vec{\theta}-\vec{\theta}_{i}\right|}.
\end{equation}

To numerically implement this expression, we evaluate Eq.~\ref{alpha_expand} over a 2D rectangular grid $\vec{\theta}_{mn}$ of $M\times N$ pixels, such that $(\Delta x_{mn,i}, \Delta y_{mn,i})$ is the displacement vector $\vec{\theta}_{mn} - \vec{\theta}_{i}$ measured from a point $m,n$ in the grid to $\vec{\theta}_{i}$, the position of the $i^{\rm th}$ galaxy. Eq.~\ref{eq:1} can then be approximated as:
\begin{equation}
\alpha_{\rm{gal}, x}\left(\vec{\theta}_{mn}\right)= \sum_{i} \alpha_{\rm gal,i} \frac{\Delta x_{mn,i}}{\left[\left(\Delta x_{mn, i}\right)^{2}+\left(\Delta y_{mn, i}\right)^{2}\right]^{1 / 2}},
\end{equation}
\begin{equation}
\alpha_{\rm{gal}, y}\left(\vec{\theta}_{mn}\right)= \sum_{i} \alpha_{\rm gal,i} \frac{\Delta y_{mn,i}}{\left[\left(\Delta x_{mn, i}\right)^{2}+\left(\Delta y_{mn, i}\right)^{2}\right]^{1 / 2}}.
\end{equation}
Here, M and N can have any arbitrary value; we use $M = N = 3000$, creating square grid of equal resolution to our GMOS-N images and with a slightly larger field of view.

Hereafter we refer to this component of the deflection field as $\vec{\alpha}\textsubscript{gal}(\vec{\theta})$. The mass distribution associated with it is given by the divergence of $\smash{\vec{\alpha}\textsubscript{gal}(\vec{\theta})}$ multiplied by half the critical surface-density (Eq.~\ref{div_alpha}). An example of the galaxy-scale mass component is shown in the top left panel of Fig.~\ref{fig:gal}. We describe in the following sections the special role of BCGs and what scaling relations \texttt{AStroLens} uses to determine the PIEMD parameters $S_{0}$,  $r\textsubscript{core}$, and $r\textsubscript{cut}$.

\subsubsection{Brightest Cluster Galaxies}
\label{sec:bcg}

Deviating again from the traditional LTM approach, \texttt{AStroLens} acknowledges the special role of BCGs in two ways. %\citep[e.g.,][and references therein]{2011MNRAS.418.2054P}.

For one, it is well known that the orientation and ellipticity of BCGs tend to align with those of the host cluster, the alignment being most pronounced for the most luminous and dominant BCGs \citep{1982A&A...107..338B,2010MNRAS.405.2023N,2019ApJ...874...84W}. Consequently, \texttt{AStroLens} uses the orientation and ellipticity of the BCG to model the cluster-scale halo (see Section~\ref{sec:clus_scale}). Since our cluster sample is diverse, comprising both dynamically relaxed systems and ongoing mergers, some clusters contain several galaxies that could be considered the BCG of the entire system or of a subcluster. If several plausible BCG candidates are present, we choose the one nearest the median luminosity-weighted position of all cluster galaxies to be the cluster's primary BCG, unless other sufficiently luminous galaxies are found within 100 kpc of it, in which case we adopt as the ellipticity and orientation angle the weighted averages (by luminosity) of the respective values for the qualifying galaxies.

In addition to using BCGs to estimate the preferential orientation and ellipticity of the large-scale mass distribution in lensing clusters, \texttt{AStroLens} also adjusts the fundamental weights used to derive the large-scale mass component to account for the presence of a special class of BCGs, encountered only in the centres of highly evolved galaxy clusters. In relaxed clusters, the BCG is often a highly luminous D galaxy with an extended low-surface-brightness halo \citep[`cD galaxy';][]{1964ApJ...140...35M,1988ApJ...328..475S}, a type of galaxy found exclusively at the bottom of the deep potential well of cluster cores \citep{,1986RvMP...58....1S,1987IAUS..127...89T,2016MNRAS.455.4442S}.

\begin{figure}
    \centering
    \includegraphics[width=\columnwidth]{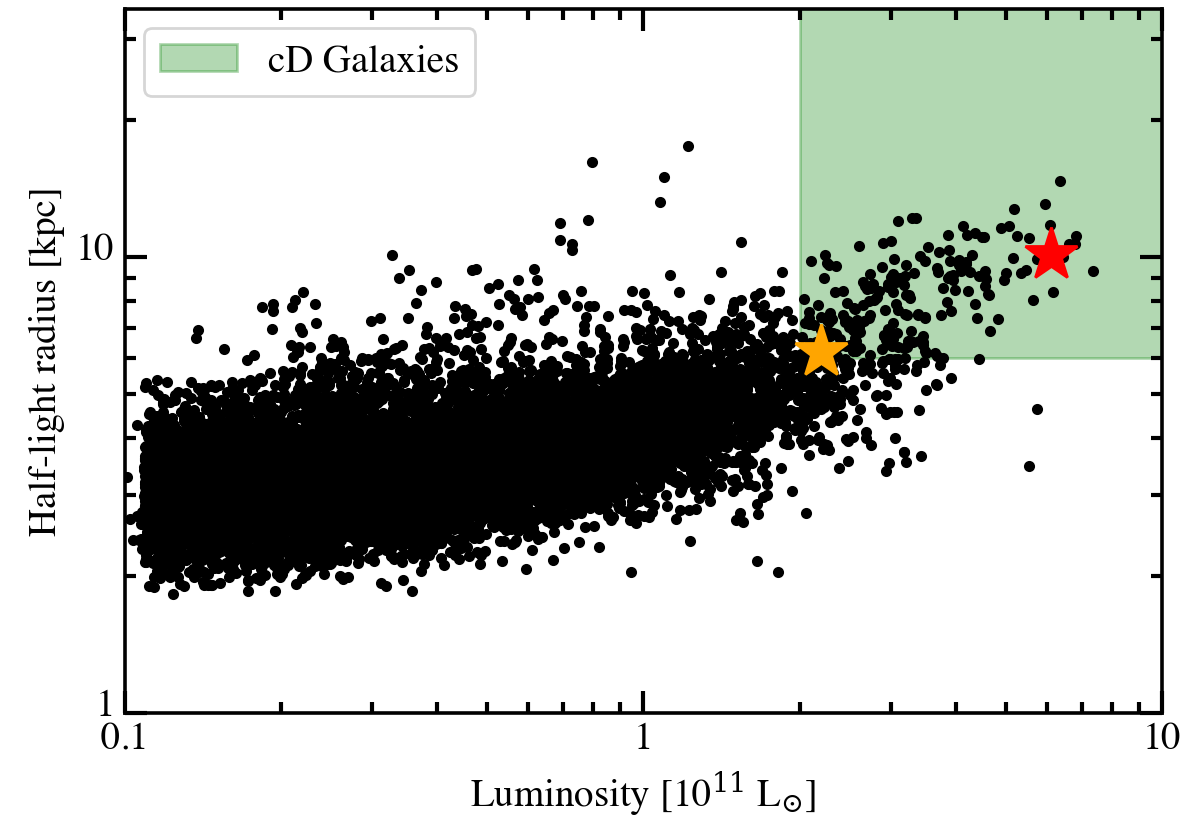}
    \includegraphics[width=0.49\columnwidth,clip,trim=17 16 14 16]{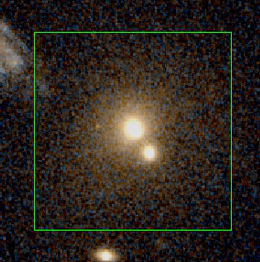}
    \includegraphics[width=0.49\columnwidth,clip,trim=22 17 17 19]{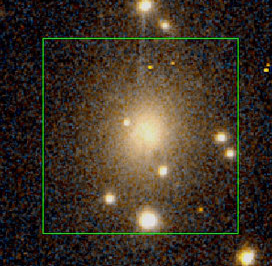}
    \caption{(Top) Distribution of luminosity (in the GMOS $i^\prime$ passband) versus half-light radius for all galaxies selected as likely members of the clusters. The box in the top right corner highlights the regime that contains the majority of the visually selected cD galaxies. (Bottom) \textit{HST} images of examples of galaxies considered marginal and \textit{bona fide} cD galaxies based on their location in the $L-r_\text{1/2}$ plane (orange and red star in diagram on top, respectively). Each image measures 50 kpc on the side at the cluster redshift.}
    \label{fig:gmos-cd}
\end{figure}

Within the local universe, groundbased imaging can detect the low-surface-brightness halos characteristic of cD galaxies to large radii; at the relatively high redshifts of our target clusters, however, this task becomes more challenging for our shallow GMOS-N imaging data. Fig.~\ref{fig:gmos-cd} highlights the realm occupied by cD galaxies in the luminosity vs half-light-radius plane of red-sequence galaxies and shows \textit{HST} images of two galaxies that we consider marginal and \textit{bona fide} cD galaxies, respectively, based on our visual inspection of BCGs in all \textsl{HST} data available for eMACS clusters (see Ebeling et al., in preparation, for details). All galaxies within the marked area (defined by $L_{{\rm i}^\prime}>2\times 10^{11} L_{\odot} \cap r_{\rm 1/2}>6$ kpc) are treated as cD galaxies by \texttt{AStroLens} and given additional weight in the light map on which the large-scale mass distribution is based (see Section~\ref{sec:clus_scale}).

\subsubsection{PIEMD scaling relations}
\label{sec:piemd}

Dedicated lens modeling packages commonly adopt the following scaling relations for the basic parameters of the PIEMD surface-brightness profile of cluster galaxies:
\begin{equation}\label{eq:scale1}
\begin{array}{l}{S_{0} = \sigma_0^2;\; \sigma_0 = \sigma_0^{\star}\left(\frac{L}{L_\text{ref}}\right)^{0.25}}\\ {r_{\text{core}}=r_{\text {core}}^{\star}\left(\frac{L}{L_\text{ref}}\right)^{0.5}}\\
{r_{\text{cut}}=r_{\text {cut}}^{\star}\left(\frac{L}{L_\text{ref}}\right)^{0.5}}\end{array}
\end{equation}
\citep[e.g.,][]{2007ApJ...668..643L}. The first of these scaling relations links the velocity dispersion of a galaxy to its luminosity and is inspired by the Faber-Jackson relation for early-type galaxies \citep{1976ApJ...204..668F}. In conjunction with the third relation, this scaling law establishes a constant mass-to-light ratio, since the total mass of a PIEMD halo scales as $S_0 r_{\rm cut}$ (Eq.~\ref{eq:piemdmass}). Extensive observational studies of the Fundamental Plane of elliptical galaxies, however, suggest that $M/L$ is not constant, but increases with galaxy luminosity as $L^{0.2-0.4}$  \citep{1996MNRAS.280..167J, 2009ApJS..182..216K}, and that $M/L$ also increases with velocity dispersion, namely as $\sigma^{0.84\pm 0.07}$ \citep{2006MNRAS.366.1126C}. Finally, the $\sigma$-$L$ relation is known to flatten at high galaxy mass, to well below the $L^{0.25}$ dependence suggested by Faber and Jackson. All of these observational constraints can be jointly met (within their uncertainties) if the luminosity dependence of the traditional scaling laws from Eq.~\ref{eq:scale1} is modified to 
\begin{equation}\label{eq:scale2}
\begin{array}{l}{\sigma_0 = \sigma_0^{\star}\left(\frac{L}{L_\text{ref}}\right)^{0.2}}\\ {r_{\text{core}}=r_{\text {core}}^{\star}\left(\frac{L}{L_\text{ref}}\right)^{0.5}}\\
{r_{\text{cut}}=r_{\text {cut}}^{\star}\left(\frac{L}{L_\text{ref}}\right)^{0.8},}\end{array}
\end{equation}
where $L_\text{ref}$ is set to $4\times10^{10}$ $L_\odot$, a value representative of the luminosity range probed by our data (see Fig.~\ref{fig:gmos-cd}), and $\sigma_0^\star$ is an \texttt{AStroLens} model parameter. The reference values $r_{\text{core}}^\star$ and $r_{\text{cut}}^\star$ are set to 0.15 and 30\,kpc, respectively. These scaling laws should hold for all galaxies, including cD galaxies, which -- although extremely luminous -- feature mass-to-light ratio consistent with that of other massive early-type galaxies \citep[e.g.,][]{1987IAUS..127...89T}. 

\subsection{Cluster-Scale Component} \label{sec:clus_scale}
In accordance with our LTM assumption, we derive the cluster-scale mass halo from the galaxy distribution by convolving the galaxy light map -- represented by luminosity-weighted Delta functions at the locations of all cluster members -- with an elliptical smoothing kernel that, again, follows a PIEMD profile. The weights of each galaxy in the construction of the light map are linear in the traditional LTM approach, i.e., each galaxy location contributes proportional to its luminosity. We modify this procedure to give extra weight to the large-scale mass distribution around all BCGs, as well as all cD galaxies (identified as shown in Fig.~\ref{fig:gmos-cd}), which are known to mark the center of a cluster's gravitational well (Section~\ref{sec:bcg}), i.e., we apply weights of
\begin{equation}
\begin{array}{ll}      
 L &\textrm{for regular cluster galaxies},\\
 L \left(\frac{L}{L_{\rm ref}} \right)^{W_L} & \textrm{for BCGs and cDs}.
\end{array}
\end{equation}
Here, $W_L$ is one of \texttt{AStroLens}' three free model parameters --- as is $R_{\rm core}$, the core radius of the large-scale PIEMD smoothing kernel\footnote{The choice for the kernel's PIEMD cut radius, set by us to 1500 kpc, does not affect the deflection field in the strong-lensing regime.}. The kernel's ellipticity and orientation angle are set to the respective characteristics of the BCG (see Section~\ref{sec:bcg}).

To summarize, our approach to modeling the large-scale mass component differs from previous work \citep{2015ApJ...801...44Z,2017ApJ...839L..11Z,2019MNRAS.482.1824S} by adopting a PIEMD smoothing kernel whose ellipticity and orientation match the respective characteristics of the BCG, and by assigning additional weight (beyond the linear luminosity weighting attached to all galaxy locations) to the BCG and all cD galaxies. An example of the large-scale mass component determined in this manner is shown in the top right panel of Fig.~\ref{fig:gal}.

Having computed the distribution of the large-scale mass component, we obtain the associated deflection field at any given pixel location by adding the contributions from all $M\times N$ pixels:
\begin{equation}
\alpha_{\rm{clus}, x}\left(\vec{\theta}_{mn}\right)\propto \sum_{i\in M,N} P_i \frac{\Delta x_{mn,i}}{\left(\Delta x_{mn,i}\right)^{2}+\left(\Delta y_{mn,i}\right)^{2}},
\end{equation}
\begin{equation}
\alpha_{\rm{clus}, y}\left(\vec{\theta}_{mn}\right)\propto \sum_{i\in M,N} P_i \frac{\Delta y_{mn,i}}{\left(\Delta x_{mn,i}\right)^{2}+\left(\Delta y_{mn,i}\right)^{2}},
\end{equation}
 where $P_{i}$ is the unnormalized surface mass density at the $i^{\rm th}$ pixel ($i\in M,N$), and constant factors have been omitted, as in Eq.~\ref{eq:1}. In practice, the above summation is carried out as a convolution in Fourier space on a grid binned 4$\times$4 and interpolated back to the original resolution. Hereafter we refer to this component of the deflection field as $\vec{\alpha}\textsubscript{clus}(\vec{\theta})$. Note that, being based solely on the observed large-scale distribution of light from cluster members, $\vec{\alpha}\textsubscript{clus}(\vec{\theta})$ still needs  normalizing (see following section).

\subsection{Total Deflection Field}
\label{sec:eta}

\begin{figure*}
	\includegraphics[width=\columnwidth]{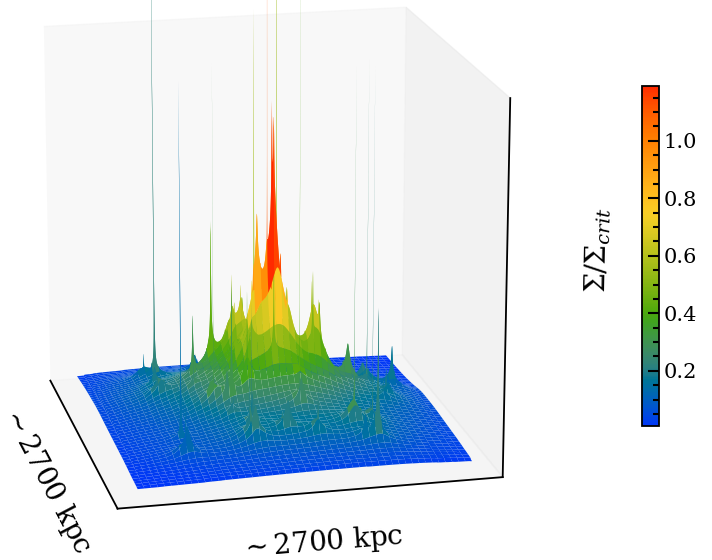}
	\includegraphics[width=\columnwidth]{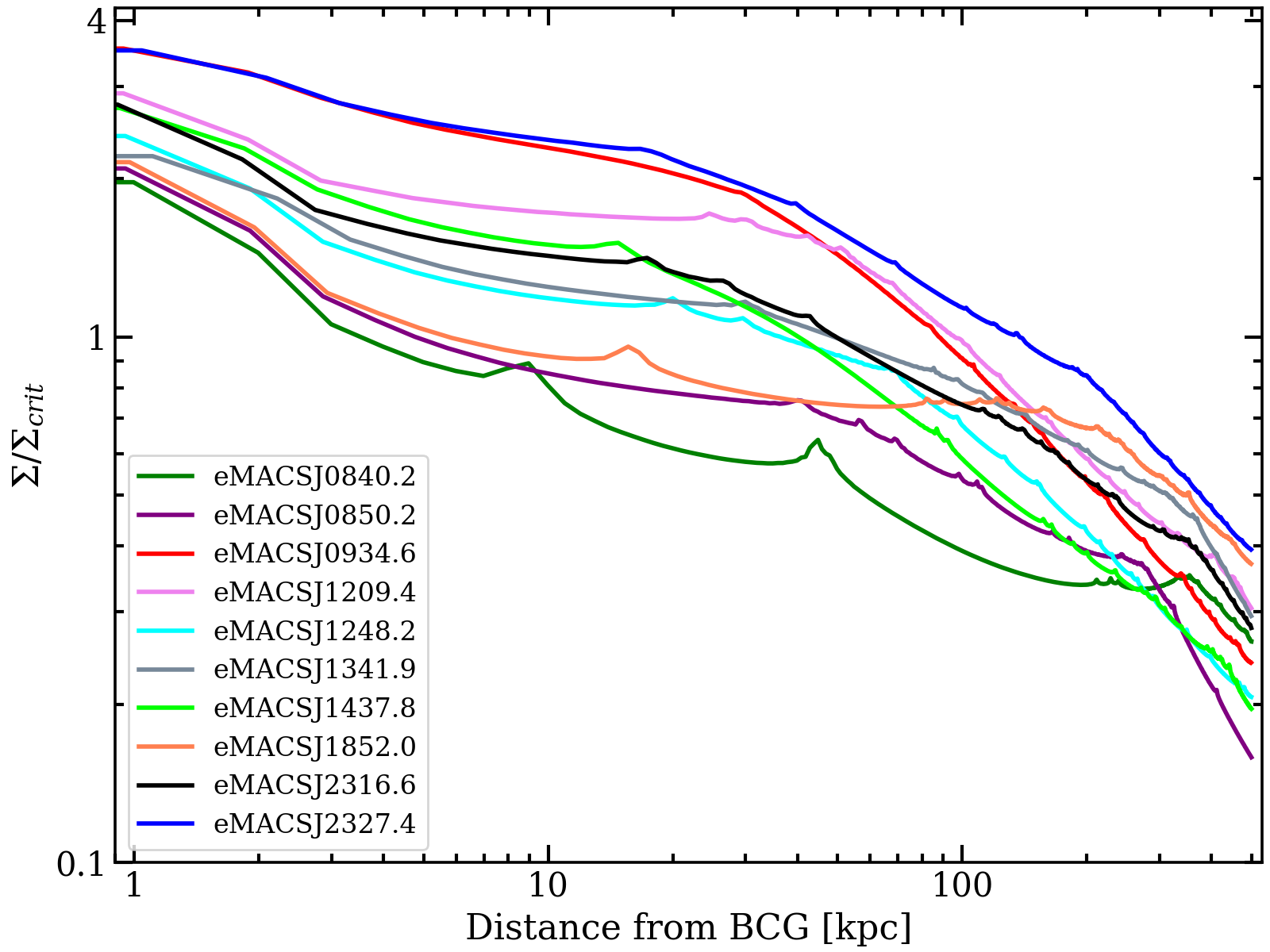}
    \caption{(Left) 3-dimensional representation of the convergence profile of eMACSJ2316 for a source at $z = 2$. (Right) Radially averaged convergence profiles for all ten clusters in the calibration sample. Each profile is centered at the location of the BCG.}
    \label{fig:e2316_3d}
\end{figure*}

The total deflection field is found by adding both components: 
\begin{equation} \label{a_tot}
\vec{\alpha}\textsubscript{T}(\vec{\theta})= \vec{\alpha}\textsubscript{gal}(\vec{\theta})+\vec{\alpha}\textsubscript{clus}(\vec{\theta}),
\end{equation}
where $\vec{\alpha}\textsubscript{clus}(\vec{\theta})$ is normalized such that $\smash[b]{\langle \alpha_\textsubscript{clus}(\vec{\theta})\rangle = \frac{1-\eta_{\rm gal}}{\eta_{\rm gal}}\langle \alpha_\textsubscript{gal}(\vec{\theta})\rangle}$. Here, angular brackets denote the average of the respective component, and $\eta\textsubscript{gal}$ quantifies the relative contribution of the galaxy-scale component\footnote{The quoted normalization requirement for $\langle \alpha_\textsubscript{clus}(\vec{\theta})\rangle$ is equivalent to $\eta_{\rm gal} = \langle \alpha_\textsubscript{gal}(\vec{\theta})\rangle / \langle \alpha_\textsubscript{T}(\vec{\theta})\rangle$, and hence $\langle \alpha_\textsubscript{T}(\vec{\theta})\rangle = \eta_{\rm gal} \langle \alpha_\textsubscript{gal}(\vec{\theta})\rangle + (1-\eta_{\rm gal}) \langle \alpha_\textsubscript{clus}(\vec{\theta})\rangle$.}. 
Having found the dependence of our results on the value of $\eta_{\rm gal}$ to be weak (we originally treated $\eta_{\rm gal}$ as an additional global parameter; see Section~\ref{sec:par_cal}), we set $\eta_{\rm gal}=0.03$, consistent with the range of 1 to 3\% reported by \citet{2010MNRAS.407..263A} for the stellar mass fraction in massive clusters. Our representation of $\vec{\alpha}\textsubscript{T}$ as a weighted sum is equivalent to the formalism used in ~\cite{2020MNRAS.491.3778C}. Fig.~\ref{fig:e2316_3d} shows, for the same cluster as in Fig~\ref{fig:gal}, a three-dimensional representation of the final mass distribution corresponding to $\vec{\alpha}\textsubscript{T}$ and radially averaged surface-density profiles for the ten calibration clusters ($z_{\rm{s}} = 2$; see Section~\ref{sec:cal_sample} for further details regarding these clusters).

%===================================================================================================
\section{Model Fitting and Parameter Calibration} \label{sec:par_cal}
As summarized in Table \ref{tab:params}, \texttt{AStroLens} uses three free parameters to derive the strong-lensing deflection field from the cluster light distribution: $\sigma_0^\star$, the velocity dispersion that determines the effective normalization of the mass profile of all cluster galaxies; $W_L$, the exponent of the special luminosity-based weight applied only to BCGs and cD galaxies; and $R\textsubscript{core}$, the core radius of the elliptical, large-scale PIEMD kernel used to smooth the galaxy light distribution.

\subsection{Minimization Formalism} \label{sec:minimization}

For \texttt{AStroLens} to credibly map the magnification and estimate the lensing strength of massive galaxy clusters, the algorithm must be able to predict the location of critical lines with reasonable accuracy. The choice of optimal values for the three global parameters used by our algorithm to derive mass models for the entire cluster sample is crucial to achieving this goal.

In order to calibrate the values of the free parameters used by \texttt{AStroLens} (see Table~\ref{tab:params}), we follow the approach of \cite{2019MNRAS.482.1824S} and require that the critical lines predicted by our code coincide with the locations of securely identified strong-lensing features observed in a subset of the clusters in our sample (described in Section~\ref{sec:cal_sample}); specifically, the predicted critical lines need to intersect known giant arcs and bisect close pairs of unambiguously identified multiple images.

To implement the required parameter optimization, we first calculate the lensing magnification on an $M\times N$ grid. The magnification $\mu$ at a position $\vec{\theta}$ in the image plane (see ~\citealt{2005ApJ...621...53B} and references therein) can be expressed through the derivatives of the deflection field
\begin{equation} \label{mu_tot}
\mu(\vec{\theta})^{-1}=1-\vec{\nabla} \cdot \vec{\alpha\textsubscript{T}}+\frac{d \alpha_{x}}{d x} \frac{d \alpha_{y}}{d y}-\left( \frac{d \alpha_{x}}{d y} \right)^2.
\end{equation}
Here $\alpha_{x}$ and $\alpha_{y}$ are the $x$ and $y$ components of the total deflection field (Eq. \ref{a_tot}). An example of a  magnification field computed in this manner is shown in  Fig.~\ref{fig:gal} (bottom).
  
Strong gravitational lensing creates both tangential and radial critical lines (schematically illustrated by the outer and inner ellipses in Fig.~\ref{fig:diagram2}, respectively). As the names imply, images formed near tangential critical lines are stretched tangentially, and images formed near radial critical lines are stretched radially. The corresponding radial ($+$) and tangential ($-$) components of the inverse magnification are given by
\begin{equation} \label{eq:lambda}
\lambda_{\pm} = 1 - \frac{\vec{\nabla} \cdot \vec{\alpha\textsubscript{T}}}{2} \pm \sqrt{\frac{1}{4}\left[\frac{d \alpha_{x}}{d x} - \frac{d \alpha_{y}}{d y}\right]^2 + \left( \frac{d \alpha_{x}}{d y}\right)^2},
\end{equation}
such that $\lambda_{+} \lambda_{-} = \mu(\vec{\theta})^{-1}$.
We here focus exclusively on constraining the location of the tangential critical lines, for the following reasons. For one, the most readily identifiable strong-lensing features, giant arcs (Fig.~\ref{fig:diagram2}), form along tangential critical lines. In addition, the location of the tangential critical line is readily identified as the mid-point of the line connecting multiple images appearing on either side. Finally, radial critical lines are confined to a very small area close to the core of the mass distribution, making radially magnified images both rare and hard to detect due to the close proximity of the BCG. 

As the magnification becomes infinite on a tangential critical line, the inverse magnification $\lambda$ approaches zero near the locations of arcs and between close pairs of multiple images. As a result, $\lambda$ has been successfully employed in prior LTM work to optimize model parameters \citep{2019MNRAS.482.1824S, 2019arXiv190509802C}. Following the same general strategy, we define a $\chi^2$ estimator for a given parameter set $P =  \{R_{\rm core}, \sigma_0^\star, W_L\}$ and a given lensing cluster as:
\begin{equation} \label{eq:chi}
\chi^2_{i}(P) =  \sum_{a} \frac{\lambda(P,\vec{\theta}_{a})^2}{\sigma_{a}^2}. 
\end{equation}

Here $\vec{\theta}_{a}$ are the locations of pixels that fall on the critical line (as estimated by us based on a known strong-lensing feature), $\lambda(P,\vec{\theta}_{a})$ is the inverse of the tangential magnification (see Eq.~\ref{eq:lambda}) evaluated at the location $\vec{\theta}_{a}$ for a set of parameters $P$, and $\sigma_{a}$ is the  uncertainty in the estimated location $\vec{\theta}_{a}$. Note that the sum in Eq.~\ref{eq:chi} runs over all selected lensing features and can thus accommodate multiple-image systems in the same cluster but at different redshifts. We account for uncertainties in our estimates for the location of the critical line by setting the uncertainty in the inverse magnification at the selected constraint location to $\sigma_a=0.1$ for all constraints\footnote{The exact value of $\sigma_a$ has little impact on the optimization process, as long as $\sigma_a$ remains below 0.3.}.

We constrain all free parameters (as part of a universal set of parameters suitable to model all clusters within our sample) by requiring that the best-fitting parameter set meet the strong-lensing constraints for several calibration clusters simultaneously, such that the total $\chi^2$, given by $\chi^2(P) = \sum_{i} \chi^2_{i}$, is minimized.

\subsection{The Calibration Sample} \label{sec:cal_sample}

Since the critical surface density and the deflection field depend on the distances between the observer and the source and between the lens and the source (Eqs.~\ref{eq:crit} and \ref{eq:alpha}), the location of critical lines can only be determined for clusters with unambiguously identified multiple-image systems that have secure redshifts, ideally from spectroscopic observations.

We use all eMACS clusters in our \textit{Gemini-N} sample that had spectroscopically confirmed multiply lensed images as of late 2019, totaling ten clusters and twelve constraints (two galaxy clusters contain two sets of multiple images)\footnote{We exclude one additional system with spectroscopic redshifts for an obvious multiple-image system on the grounds of the deflection being primarily due to galaxy-galaxy lensing.}. As these systems exhibit a variety of lensing configurations and nearly span the full redshift range of our sample, we expect the calibration set to be representative of the entire eMACS sample. We note that, in some cases, the relevant strong-lensing features were initially identified in \textit{HST} images (see Ebeling et al., in preparation, for details) and subsequently merely rediscovered in our ground-based data.

As part of the first public release of the eMACS cluster sample we list all calibration clusters, together with basic physical properties, in Table~\ref{tab:emacs-cal}. Below we briefly describe notable characteristics and the constraints provided by the ten calibration clusters.

\begin{table*}
    \centering
    \begin{tabular}{ccccccc}
    name & R.A.\ (J2000) Dec.\ & $z$ & $n_z$ & $L_{\rm X}$ & $\theta_{\rm E}$ & $z$ reference\\ & & & & [$10^{44}$ erg s$^{-1}$] & [\arcsec] & \\[2mm] \hline
        eMACSJ0840.2$+$4421 & 08 40 09.4  \,\,$+$44 21 54 &0.639 & 30 & $14.1\pm 4.3$\,\,\, &13 & \citet{2013MNRAS.432...62E}\\
        eMACSJ0850.2$-$0611 & 08 50 12.9  \,\,$-$06 12 22 &0.574 & 22 & $5.4\pm 2.1$ &16 & \\
        eMACSJ0934.6$+$0540 &  09 34 39.0 \,\,$+$05 41 47 &0.561 & 6 & $10.1\pm 2.0$\,\,\, &28 & SDSS\\
        eMACSJ1209.4$+$2640 & 12 09 23.7  \,\,$+$26 40 47 &0.558 & 18 & $6.1\pm 1.7$ &28 & NED\\
        eMACSJ1248.2$+$0743 & 12 48 16.7  \,\,$+$07 42 58 &0.573 & 15 & $5.2\pm 1.9$ &15 & \\
        eMACSJ1341.9$-$2442 & 13 42 00.2  \,\,$-$24 42 01 &0.834 & 26& $16.2\pm 6.2$\,\,\, &22 & \citet{2018ApJ...852L...7E}\\
        eMACSJ1437.8$+$0616 & 14 37 49.8  \,\,$+$06 16 41 &0.535 & 19 & $5.8\pm 2.1$ &11 & \\
        eMACSJ1852.0$+$4900 & 18 52 02.7  \,\,$+$49 01 18 &0.604 & 32 & $6.3\pm 1.7$ &21 & \\
        eMACSJ2316.6$+$1246 & 23 16 42.9  \,\,$+$12 46 55 &0.526 & 32 & $4.9\pm 1.6$ &17 &\\
        eMACSJ2327.4$-$0204 & 23 27 27.6  \,\,$-$02 04 37 &0.706 & 24 & $14.0\pm 4.1\,\,\,$ &40 & \\
         & 
    \end{tabular}
    \caption{The ten eMACS clusters used to calibrate the three global model parameters of \texttt{AStroLens}. eMACSJ2327 is also one of the ten strongest eMACS cluster lenses (Table~\ref{tab:emacs}). The listed coordinates correspond to the location of the BCG(s); X-ray luminosities are derived from RASS data in the 0.1--2.4 keV band; Einstein radius estimates assume a fiducial source redshift of $z=2$ (see Section~\ref{sec:erads} for details). Redshifts were measured by the eMACS team, unless stated otherwise.}
    \label{tab:emacs-cal}
\end{table*}

\begin{description}
\item[\bf eMACSJ0840.2$+$4421] was discovered in the eMACS pilot programme \citep{2013MNRAS.432...62E} and features a massive cD galaxy. In spite of a bright star nearby, a giant arc consisting of three individual images %($z = 2.23$) 
stands out near the cluster centre. 
\item[\bf eMACSJ0850.2$-$0611] shows several obvious strong-lensing features; at this time, the only one with a secure spectroscopic redshift is a fold arc close to the second-brightest cluster galaxy. 
\item[\bf eMACSJ0934.6$+$0540] is home to a dual BCG and several multiple-image systems. We use a fold arc at the SW end of the cluster core for calibration purposes. 
\item[\bf eMACSJ1209.4$+$2640] was previously discovered in the course of the Sloan Giant Arc Survey and the subject of a dedicated modeling effort  \citep{2020ApJS..247...12S}. 
The giant arc east of the BCG is the lensed image of a galaxy at $z = 1.02$.
\item[\bf eMACSJ1248.2$+$0743] shows several obvious strong-lensing features in an archival \textsl{HST} SNAPshot. However, at present only one triple image is spectroscopically confirmed. %($z = 1.27$).
\item[\bf eMACSJ1341.9$-$2442] is one of the most distant clusters in our sample and known for its extreme magnification of a quiescent galaxy at $z = 1.59$  \citep{2018ApJ...852L...7E}. We adopt the midpoint between the two eastern images of the triple as the location of the corresponding critical line.
\item[\bf eMACSJ1437.8$+$0616] features two giant arcs on either side of the BCG that were originally discovered in WFC3/IR images. Both arcs were found to be at the same redshift % $z \sim 2.55$ 
and, as distinct systems, yield an independent constraint each for the minimization described in Section~\ref{sec:minimization}.
\item[\bf eMACSJ1852.0$+$4900] is a highly elongated system with a string of galaxies of similar brightness at its centre. The associated mass distribution lenses a galaxy at $z = 1.54$ into a nearly straight fold-arc configuration of three images. 
\item[\bf eMACSJ2316.6$+$1246] is the eMACS cluster used as an example of the modeling procedure adopted by \texttt{AStroLens}, e.g., in Figs.~\ref{fig:gal} and \ref{fig:e2316_3d}. Its dual BCGs are accompanied by two spectroscopically confirmed multiple-image families %, at $z = 1.46$ and $z = 2.46$, respectively. 
The first one appears as a giant, almost straight arc between the two BCGs, while the second is a complex fold arc that crosses the critical lines near the eastern BCG.  
\item[\bf eMACSJ2327.4$-$0204] was discovered previously in the Red-sequence Cluster Survey 2 (RCS2,~\citealt{2011AJ....141...94G}) and is one the most massive galaxy clusters known at $z\sim 0.7$ ~\citep{2015ApJ...814...21S}. Many lensed images in the central region can be discerned, even in the GMOS-N data. However, we use only one multiple-image system for our calibration: a galaxy $z = 2.98$ lensed into a cusp configuration far from the cluster core. 
\end{description}

\begin{figure*}
    \centering
        \includegraphics[width=\textwidth]{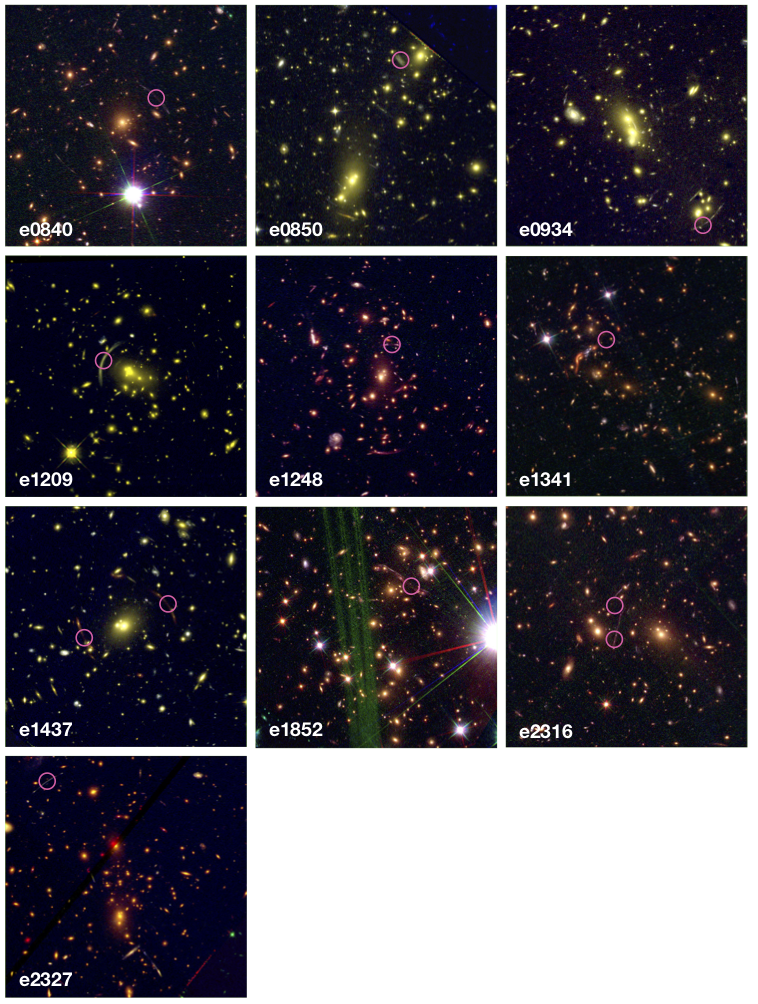}\\[-50mm]
        \hspace*{55mm}\parbox{0.6\textwidth}{\caption{\textsl{HST} colour images (both ACS and WFC3 data are used) of the ten \texttt{AStroLens} calibration clusters. All images span 1.5 arcmin on a side; North is up and East is to the left. Magenta circles mark the approximate locations of the critical line near multiple images used in the \texttt{AStroLens} calibration process. Note that both eMACSJ1437.8 and eMACSJ2316.6 contain two sets of spectroscopically confirmed multiple images and hence provide two constraints each.\label{fig:hst_cal}
        }}\mbox{}\\[30mm]
        
\end{figure*}

The pixels selected to specify the locations of critical lines, $\vec{\theta}_{a}$, lie typically close to the centres of giant arcs or close to the midpoint between the two multiple images nearest each other for a fold arc (Fig.~\ref{fig:diagram2}). The resulting model constraints are weighted as described in Section~\ref{sec:minimization}. We show \textsl{HST} images of the relevant region of each calibration cluster in Fig.~\ref{fig:hst_cal} as well as the locations $\vec{\theta}_{a}$ used during the minimization.

\subsection{MCMC Minimization}\label{sec:mcmc}

The parameter values that minimize $\chi^2$ as defined in Eq.~\ref{eq:chi} are found through a standard Markov Chain Monte Carlo (MCMC) maximum-likelihood process, where the likelihood is defined as $L = e^{-\chi^{2}/2}$. In practice, we employ the affine invariant MCMC sampler \texttt{emcee} \citep{2013PASP..125..306F}. Unlike a single Markov chain, which is susceptible to getting trapped in a local $\chi^{2}$ minimum, \texttt{emcee} uses an ensemble of so-called walkers that explore the parameter space widely and simultaneously. A single walker is similar to a Markov chain except that its proposal distribution depends on the current positions of all other ensemble walkers. By design \texttt{emcee} thus takes advantage of parallel processing to find and explore the parameter volume of highest likelihood.

Prior to the MCMC iteration process, we manually set all parameters to initial values that produce broadly acceptable critical lines. We then proceed with parameter sampling two phases. 

\begin{figure}
	\includegraphics[width=\columnwidth]{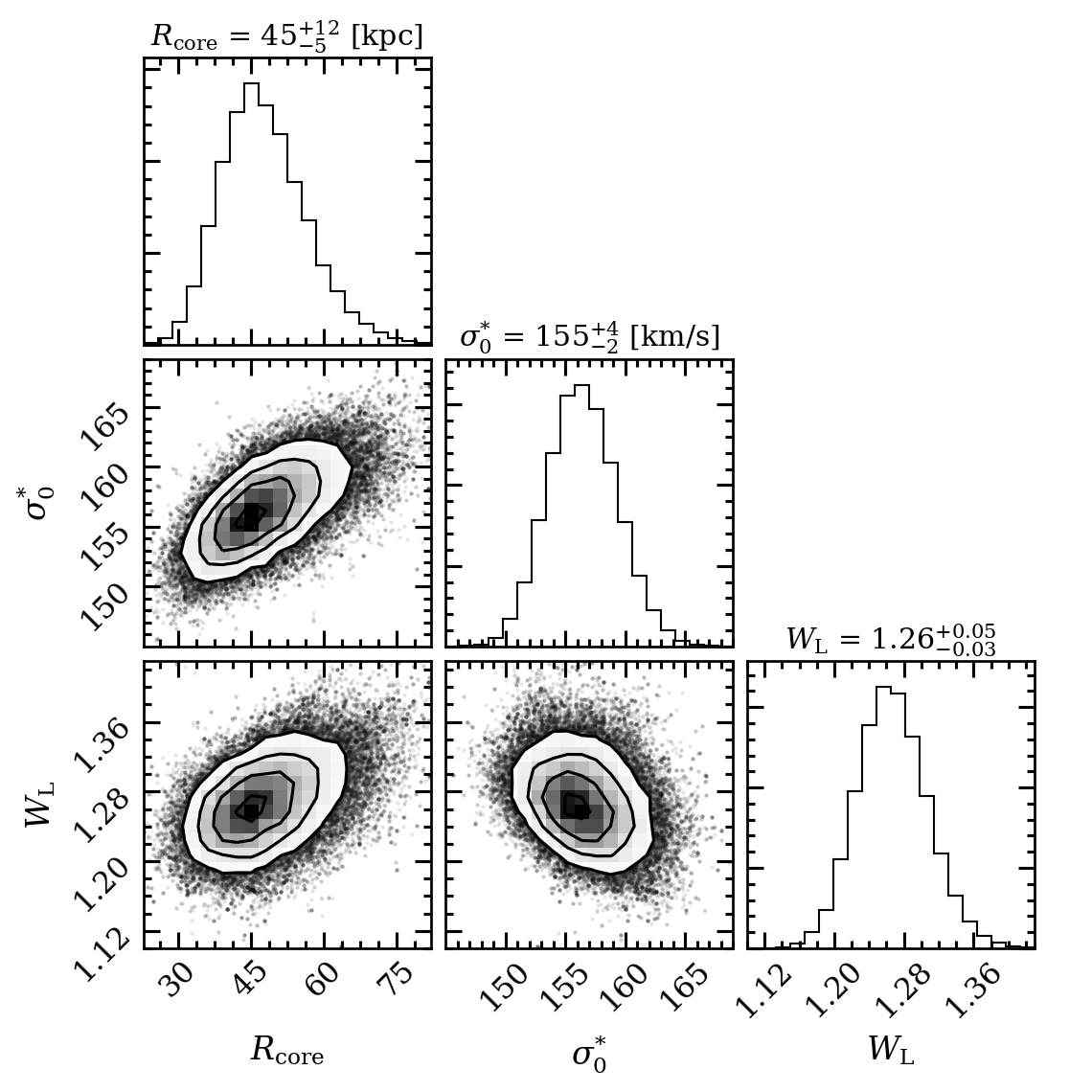}
    \caption{Posterior probability distributions and covariances for our final set of free parameters, as obtained from the analysis of our ten calibration clusters at low resolution. The values cited above each histogram correspond to the mode and the 16th and 84th percentiles of the respective distribution. (Note that the parameter values yielding the highest likelihood in our high-resolution analysis differ slightly from the locations of the peaks of the histograms shown here.)}
    \label{fig:corner}
\end{figure}

In the first phase, we work with $3\times3$ binned images for all calculations, which dramatically increases computing speed (in particular for the numerous convolutions) and allows us to efficiently explore a wide parameter space with 100,000 total MCMC steps. Fig.~\ref{fig:corner} displays the result of this step as a standard ``corner plot", highlighting both the marginalized posterior probability distributions function (PDF) for each parameter as well as the covariances between the parameters. As an estimate of the best-fit parameter values we quote the mode of each PDF, with uncertainties derived from the 16th and 84th percentiles of the distributions.

We use the results from the MCMC run described above for the second phase of our sampling procedure, which is conducted on the full $3000\times3000$ pixel grid. Each walker is initialized at parameter values drawn from Gaussian random distributions centred on the optimal values derived in the first phase. Since these parameter values are already well constrained, this second phase ensures that the optimal solution is found at the higher resolution, accounting for any discrepancies that arise from performing calculations at different resolutions. We note that this approach provides more accurate transitions between grid resolutions than the standard practice cited in the literature of simply interpolating from a lower- to a higher-resolution grid. 

Initial investigations into the \texttt{AStroLens} parameter space using this approach showed some parameters to be correlated or poorly constrained. An example is $\eta_{\rm gal}$, the relative weight of the galaxy-scale contribution to the total deflection field, which has little impact on the performance on our algorithm (see also Section~\ref{sec:eta}). We set $\eta_{\rm gal}=0.03$, consistent with the expectation that the galaxy-scale component amounts to only a small fraction of the total mass of a galaxy cluster, the majority of which is dominated by the smooth, cluster-scale component. Indeed, \citet{2018arXiv180703793C}  showed (although for a slightly different modeling scheme and goals) that an $\eta_{\rm gal}$ value of 0.1 is still suitable to describe lenses of various scales, from groups to massive clusters.

\begin{table*}
\caption{\texttt{AStroLens} parameters: description, initial value, priors, and MCMC-optimized final value. Note that the final values reflect the results from the high-resolution MCMC.}
\centering
    \begin{tabular}{llcccc} % Column formatting, @{} suppresses leading/trailing space
       \multicolumn{4}{l}{} \\
       \hline 
       \hline
        & Description & Lower Limit & Initial Value & Upper Limit & Final value \\
       \hline
       $\sigma_0^\star$ [km s$^{-1}$]   & Normalization of galaxy surface-density profile & 100    & 160 &  250 & 163  \\
       $W_L$   & Exponent of luminosity weight for BCGs &0.5    & 1 &  3 & 1.28   \\
       $R\textsubscript{core}$ [kpc]   & Core radius of elliptical PIEMD smoothing kernel & 10    & 30 &  120 & 48  \\
       \hline
    \end{tabular}
    \label{tab:params}
\end{table*}

For the three parameters $R_{\rm core}$, $\sigma_0^\star$, and $W_L$ that need to be calibrated, we perform the described minimization procedure on the High Performance Computer (HPC) cluster managed by the University of Hawai'i Information Technology Services department, starting from the initial parameter values listed in Table~\ref{tab:params}. The final MCMC sampling uses between one and two dozen processors, each equipped with 10Gb of RAM. 

Based on the process described above, we adopt as our global \texttt{AStroLens} parameters the values that minimize $\chi^2$ as defined in Eq.~\ref{eq:chi}: $R_{\rm{core}} = 48$ kpc, $\sigma_0^{\star} = 163 $ km s$^{-1}$, and $W_L = 1.28$. 

\begin{figure*}
\centering
    \includegraphics[width=\textwidth]{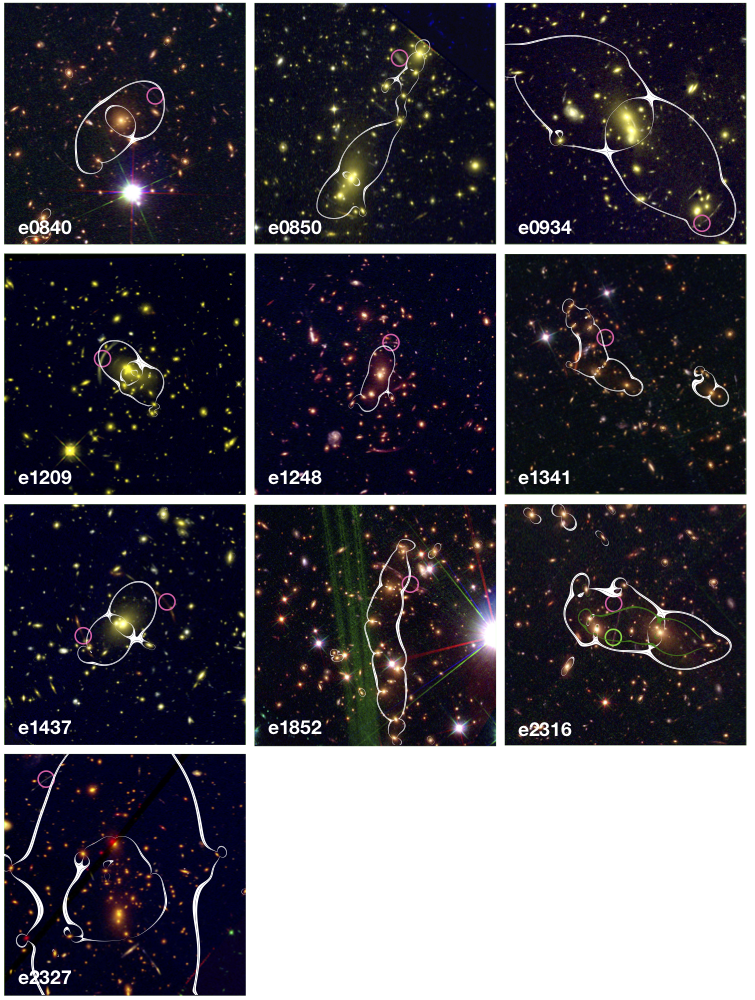}\\[-50mm]
        \hspace*{55mm}\parbox{0.6\textwidth}{
    \caption{As Fig.~\ref{fig:hst_cal}, but also showing the critical lines (white) as determined by \texttt{AStroLens} for the spectroscopically measured source redshifts and using the MCMC-optimized global parameter set. (For eMACSJ2316 we show the second strong-lensing constraint and the associated critical line in green.)}
    }\mbox{}\\[30mm]
    \label{fig:hst_cal_CL}
\end{figure*}

%===================================================================================================
\section{Results} \label{results}

We here present the results obtained by \texttt{AStroLens} for our sample of eMACS clusters, using the global parameter set calibrated as described in Section~\ref{sec:par_cal} and listed in Table~\ref{tab:params}.

\subsection{eMACS calibration clusters}

Figure~\ref{fig:hst_cal_CL} shows the critical lines predicted by \texttt{AStroLens} for the ten calibration clusters. As is to be expected for this diverse cluster sample, the critical lines do not always pass exactly through the locations defined by the provided strong-lensing constraints; the RMS distance from the predicted critical lines to the specified locations $\vec{\theta}_{a}$ is very modest though, at 2.1\arcsec\ (mean distance: 1.8\arcsec). 

\subsection{Comparison with state-of-the-art lens models}

\begin{figure}
	\hspace*{12mm}\includegraphics[width=0.7\columnwidth]{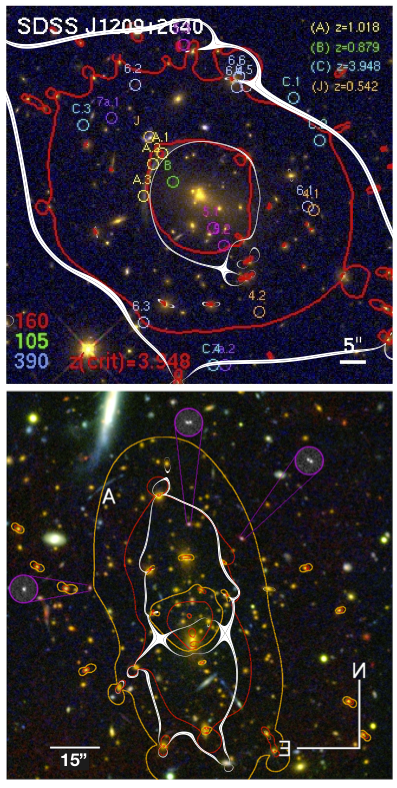}
    \caption{Top: Critical lines (shown in red) of eMACSJ1209.4 for a source at $z = 3.95$ derived by \citet{2020ApJS..247...12S} from a state-of-the-art strong-lensing model of this cluster. The \textsl{HST} filters used to create the image are listed in the lower left corner, and the redshifts of several multiple-image families (indicated by coloured circles) are given in the upper right corner. We overlay the critical line predicted by \texttt{AStroLens} in white. Figure reproduced and adapted from \citet{2020ApJS..247...12S}. Bottom: Strong-lensing analysis of eMACSJ2327.4 as performed by \citet{2015ApJ...814...21S}; the red and orange critical lines are computed for the redshifts of 1.42 and 2.98, respectively, of the two spectroscopically confirmed multiple-image families in this system. Shown in white is the critical line for a source at $z=1.42$ predicted by \texttt{AStroLens}. Figure reproduced and adapted from \citet{2015ApJ...814...21S}. 
    }
    \label{fig:lit_comp}
\end{figure}

To test the reliability of our LTM algorithm, we compare the critical lines predicted by \texttt{AStroLens} with those from previous, detailed lens modeling work for strong-lensing features not used in our calibration procedure (Section~\ref{sec:par_cal}).

\subsubsection{eMACSJ1209.4}

\citet{2020ApJS..247...12S} developed a dedicated mass model for this cluster (SDSS\,J1209+2640), using over 20 spectroscopically confirmed strong-lensing features as constraints. With the help of high-quality \textsl{HST} observations and the lens-modeling software \texttt{lenstool} ~\citep{2007NJPh....9..447J}, Sharon and coworkers produced a detailed model capable of reproducing the locations of the lensed images to high accuracy. We show the critical lines predicted by their model for one of the multiple-image systems ($z = 3.948$) in the top panel of Fig.~\ref{fig:lit_comp}, along with the \texttt{AStroLens} prediction for this source redshift. The effective Einstein radius of 22.7$^{\prime\prime}$ reported by Sharon and colleagues for a fiducial source at $z=2$ agrees to within 15\% with our value of 26.4$^{\prime\prime}$. Note that the calibration of \texttt{AStroLens}' global model parameters used the location of the central giant arc ($z = 1.02$) in this cluster, but not the multiple-image family at $z = 3.948$ for which critical lines are compared in Fig.~\ref{fig:lit_comp} (top). 

\subsubsection{eMACSJ2327.4}

This cluster was previously discovered during the Red-Sequence Cluster survey as RCS2\,J232727.6$-$020437. An extensive analysis of the system's properties, performed by \citet{2015ApJ...814...21S}, includes a dedicated strong-lensing analysis anchored by two multiple-image systems with spectroscopic redshifts of $z=1.42$ and $z=2.98$. We used the latter (the giant arc visible in Fig.~\ref{fig:hst_cal}) during the calibration of \texttt{AStroLens}' global parameters, but now focus on the second multiple-image system in eMACSJ2327.4.
The bottom panel of Fig.~\ref{fig:lit_comp} shows in red the critical line computed by \citet{2015ApJ...814...21S} for a source at $z=1.42$ (the associated images are marked by purple circles); the white line shows our \texttt{AStroLens} prediction for the same source redshift. The corresponding effective Einstein radii are 25.9$^{\prime\prime}$ \citep{2015ApJ...814...21S} vs 24.8$^{\prime\prime}$ (\texttt{AStroLens}). Again, the excellent agreement (better than 5\%) is remarkable, given that \texttt{AStroLens} is completely ignorant of the strong-lensing constraints provided by this triple image.

\subsection{Lensing strengths of eMACS clusters}

With the best-fitting values for the model parameters determined, we use \texttt{AStroLens} to model the full sample of 96 eMACS clusters with Gemini imaging data as of 2019. In the following we present the resulting effective Einstein radii, show the critical lines of the most powerful lenses identified by \texttt{AStroLens}, and discuss systems that, based on their very small estimated Einstein radii, appear not to be clusters at all. For ease of comparison (and since the redshifts of most strong-lensing features in eMACS clusters are yet to be measured), all relevant quantities are calculated for a source at the same fiducial redshift of $z = 2$ as used throughout this paper.

\subsubsection{Einstein Radii} \label{sec:erads}

\begin{figure}
	\includegraphics[width=\columnwidth]{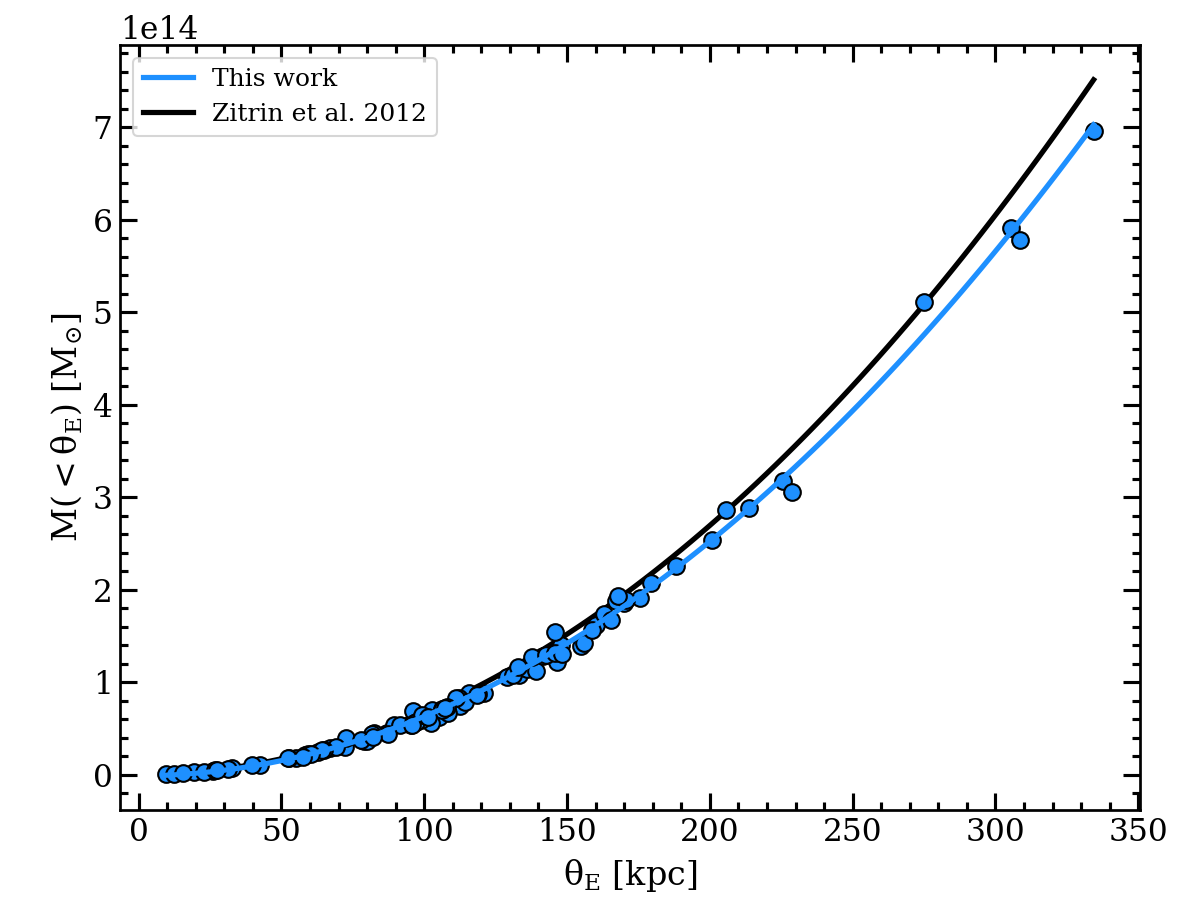}
    \caption{Einstein mass vs Einstein radius for all 96 eMACS clusters in our sample. The best-fit quadratic relation agrees well with the equivalent relation from \citet{2012MNRAS.423.2308Z}. The shown scatter is caused by the asymmetry of the mass distribution of many lenses.}
    \label{fig:erad_emass}
\end{figure}

We measure the effective Einstein radius of each cluster (see Eq.~\ref{eq:Einstein Radius}) as $\theta\textsubscript{E} = \sqrt{A/\pi}$ where $A$ is the area enclosed by the tangential critical line. Since the strong-lensing regime is non-contiguous for some clusters\footnote{An example is eMACSJ1341.9 which, at least for a source at $z=1.49$, consists of two disjoint regions (see Fig.~\ref{fig:hst_cal_CL}).}, $A$ is taken to be the sum of the three largest areas for each cluster. As a consistency check, we plot in Fig.~\ref{fig:erad_emass} the mass within the critical area as a function of the measured Einstein radius and find a quadratic relation, as expected (see Section~\ref{sec:overview}) and in excellent agreement with the best-fit polynomial found in the modeling of 10,000 SDSS clusters by ~\citet{2012MNRAS.423.2308Z}.

\begin{figure}
	\includegraphics[width=\columnwidth]{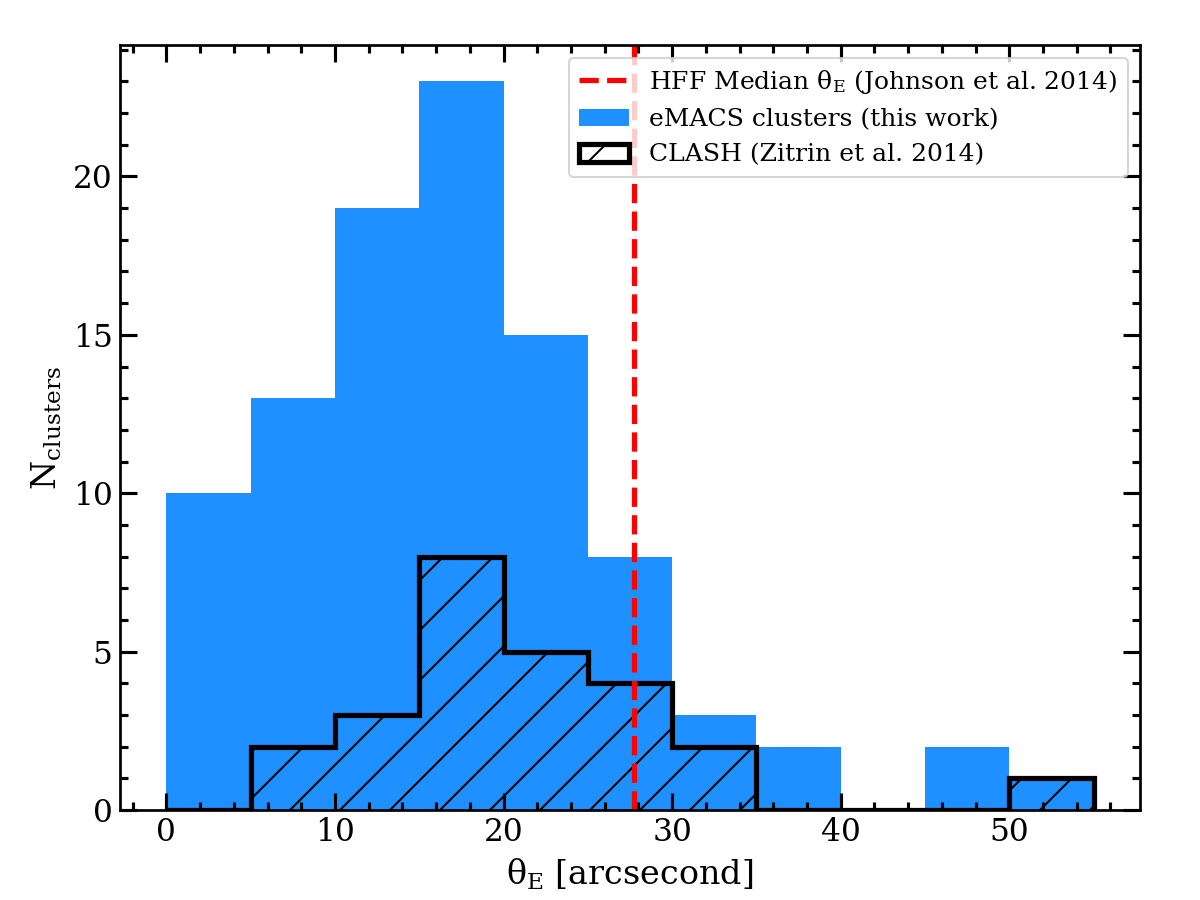}
    \caption{Distribution of effective Einstein radii measured with \texttt{AStroLens} for eMACS clusters. Also shown (hatched) is the equivalent distribution for the cluster sample observed for the CLASH project and (dashed) the median effective Einstein radius of the clusters observed during the Hubble Frontier Fields initiative.}
    \label{fig:erad_hist}
\end{figure}

The distribution of effective Einstein radii for all 96 clusters in our sample is shown in Fig.~\ref{fig:erad_hist}. We find a median Einstein radius of $\theta\textsubscript{E} = 15.9$\arcsec. Most notably, we identify 31 lenses with $\theta\textsubscript{E} \geq $ 20\arcsec, 16 lenses with $\theta\textsubscript{E} \geq $ 25\arcsec, and eight lenses with $\theta\textsubscript{E} \geq $ 30\arcsec. For comparison, we also show the distribution of Einstein radii of the 25 cluster lenses selected for the CLASH survey \citep{2012ApJS..199...25P} in Fig.~\ref{fig:erad_hist}. To put these numbers into perspective, note that four of the six massive clusters selected for the Hubble Frontier Fields initiative have Einstein radii in the range 20\arcsec $< \theta\textsubscript{E} < $ 30\arcsec\ and only two feature   $\theta\textsubscript{E} > 35$\arcsec\ ~\citep{2014ApJ...797...48J} for a source at $z = 2$. Furthermore, the largest lenses modeled by the RELICS team have Einstein radii in a similar range, with very few lenses at $\theta\textsubscript{E} > $ 30"
~\citep{2018ApJ...859..159C,2018ApJ...858...42A,2018ApJ...863..145C}.

\begin{figure*}
    \centering
        \includegraphics[width=\textwidth]{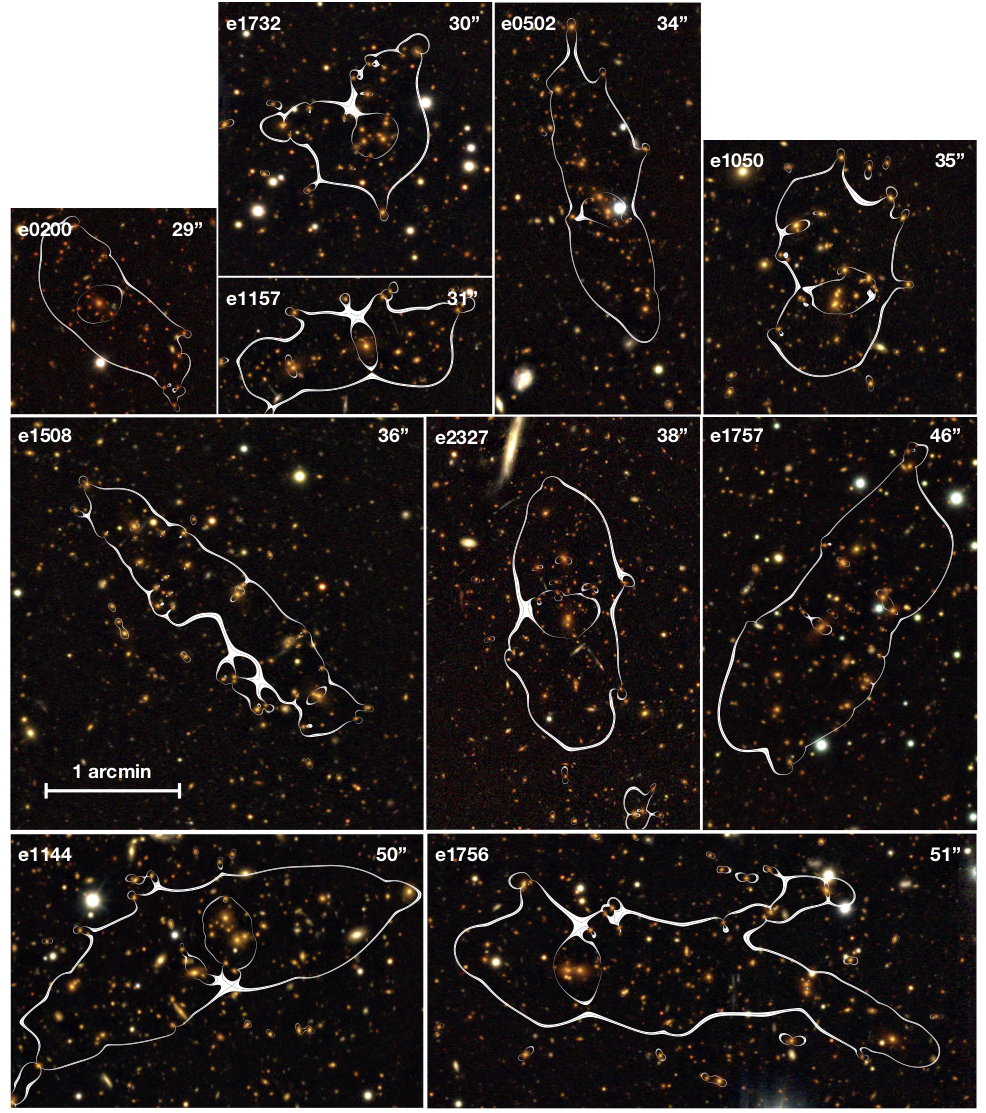}
        \caption{Critical lines (for a putative source at $z = 2$) predicted by \texttt{AStroLens} overlaid on our \textsl{Gemini-N} images of the ten eMACS clusters from our sample with the largest effective Einstein-radius estimates (listed in the top right corner of each panel). All images are shown on the same angular scale. Basic physical properties of all shown clusters are listed in Table~\ref{tab:emacs}.}
        \label{fig:result_clus}
\end{figure*}

\subsubsection{The ten most powerful eMACS lenses}

The lensing strength estimates provided by \texttt{AStroLens} allow the selection of the most promising targets for in-depth study and evaluation as powerful natural telescopes. Fig.~\ref{fig:result_clus} shows the Gemini-N images of the ten eMACS clusters with the largest effective Einstein radii in our sample, as well as the critical lines predicted by \texttt{AStroLens}. Since we again compute the critical lines for a putative source at $z = 2$, obvious strong-lensing features discernible already in our shallow groundbased images are not necessarily expected to fall on the shown critical lines.

We list these ten clusters and their basic physical properties in Table~\ref{tab:emacs} as part of the first installment of the public release of the eMACS cluster sample. We refer to \citet{2013MNRAS.432...62E} and Ebeling et al.\ (in preparation) for a comprehensive review of the eMACS project and to forthcoming papers for results of ongoing multi-wavelength analyses of the properties of individual clusters.

\begin{table*}
    \centering
    \begin{tabular}{ccccccc}
    name & R.A.\ (J2000) Dec.\ & $z$ & $n_z$ & $L_{\rm X}$ & $\theta_{\rm E}$ & $z$ reference\\ & & & & [$10^{44}$ erg s$^{-1}$] & [\arcsec] &\\[2mm] \hline
        eMACSJ0200.3$-$2454 & 02 00 16.3  \,\,$-$24 54 52 &0.713 & 25 & $8.5\pm 2.5$ &29 &\\
        eMACSJ0502.9$-$2902 &  05 02 54.6  \,\,$-$29 02 21 &0.604 & 19 & $6.7\pm 1.9$ &34 &\\
        eMACSJ1050.6$+$3548 & 10 50 38.6  \,\,$+$35 49 13 &0.508 & 26 & $4.3\pm 1.6$ &35 &\\
        eMACSJ1144.2$-$2836 & 11 44 09.5  \,\,$-$28 35 00 &0.507 & 23 & $16.5\pm 5.2$\,\,\, &50 &\\
        eMACSJ1157.9$-$1046 &  11 57 57.3  \,\,$-$10 46 01 &0.557 & 25 & $7.6\pm 3.5$ &31 &\\
        eMACSJ1508.1$+$5755 & 15 08 17.1  \,\,$+$57 54 37&0.539 & \,\,\,4& $7.2\pm 1.5$ &36 & SDSS\\
        eMACSJ1732.4$+$1934 & 17 32 24.1  \,\,$+$19 33 17 &0.541 & 24 & $4.8\pm 1.6$ &30 &\\
        eMACSJ1756.8$+$4008 & 17 56 52.6  \,\,$+$40 08 07 &0.574 & 54 & $8.6\pm 1.8$ &51 &\\
        eMACSJ1757.5$+$3045 & 17 57 29.4  \,\,$+$30 45 54 &0.611 & 32 & $6.1\pm 1.7$ &46 &\\
        eMACSJ2327.4$-$0204 & 23 27 27.6  \,\,$-$02 04 37 &0.706& 24 & $14.0\pm 4.1\,\,\,$ &38 &\\
         & 
    \end{tabular}
    \caption{The ten eMACS clusters from our sample with the largest effective Einstein radii, according to \texttt{AStroLens}. Columns as described in Table~\ref{tab:emacs-cal}. eMACSJ2327 is also a calibration cluster for this work.}
    \label{tab:emacs}
\end{table*}

\subsubsection{Potential misidentifications}

In addition to many systems with impressively large effective Einstein radii, Fig.~\ref{fig:erad_hist} also shows a substantial number of fields with very small critical areas; for five of them, $\theta_{\rm E}$ does not even exceed 4\arcsec. Since eMACS is an X-ray selected sample, such low lensing strength suggests that that much, or all, of the X-ray emission recorded in the respective direction during the RASS has a non-cluster origin. Because none of these objects has since been observed with X-ray facilities capable of yielding better photons statistics and higher spatial resolution, a credible alternative identification remains, however, in most cases elusive.

\section{Uncertainties} \label{sec:uncertainties}

Relying on photometry for a user-selected set of galaxies presumed to be cluster members, and resting on the LTM paradigm's substantial simplifications, \texttt{AStroLens}' performance is affected by both statistical and systematic uncertainties.
We here list and briefly discuss the primary sources of error identified by us:

\begin{description}
\item[\textbf{Observation design}] Since spectroscopic redshifts are rarely avai\-lable for all galaxies within the area of investigation (and down to the aimed-at magnitude limit), photometric redshifts (including red-sequence colours) derived from multi-passband photometry are key to the selection of likely cluster members. Specifically, the most prominent spectroscopic feature of late-type cluster galaxies, the 4000\AA\ continuum break, is most powerful as a redshift diagnostic when bracketed by the chosen filters  (Fig.~\ref{fig:kcorrect}). In addition, the imaged solid angle must be sufficient to cover the cluster extent, as a limited field of view may cause part of the strong-lensing regime to remain unmapped, leading to curtailed critical lines and potentially severely underestimated lensing strengths.
\item[\textbf{Photometry}] We use this term to represent all aspects of the quality of the imaging data from which the catalogue of cluster members is derived. Low angular resolution due to poor seeing increases the probability of blending in dense cluster cores (which in turn leads to erroneous photometry) and also limits our ability to faithfully determine the ellipticity and orientation of BCGs, both of which are important for the modeling of the cluster-scale mass distribution (Section~\ref{sec:clus_scale}). Extinction as the result of non-photometric conditions causes shifts in the observed colours of galaxies, complicating the selection of likely cluster members from the red sequence. And, finally, the cosmetic quality of the input images is important on all scales -- fringe and illumination corrections, ghosts, stellar diffraction spikes, and other artifacts all affect the observer's ability to compile a clean source catalogue and obtain robust photometry and hence representative light maps.
\item[\textbf{Galaxy selection}] Since \texttt{AStroLens} is designed to model the deflection field due to strong lensing for many clusters simultaneously, the galaxies selected to represent the light distribution within these clusters must be selected homogeneously and consistently across the sample. Specifically, photometric redshifts (or red-sequence membership) ought to be established in a self-consistent manner for all clusters, and presumed cluster members must be selected within the same absolute-magnitude (luminosity) range (Section~\ref{sec:rs_select}). 
\item[\textbf{Model calibration}]
A significant number of constraints, preferably involving a range of lensing configurations and source redshifts, is essential for a robust calibration and \texttt{AStroLens} results that are credible for a similarly diverse set of non-calibration clusters. Since the calibration of \texttt{AStroLens}' global parameters is critical for the performance of the algorithm, any uncertainties in the exact location (both in the plane of the sky and in redshift space) of strong-lensing features used for calibration purposes (see Section~\ref{sec:par_cal}) should be minimized and quantified. 
\item[\textbf{Collisional matter}] The fundamental observable at the heart of the LTM paradigm is the light from cluster galaxies. Since the distribution of cluster galaxies is strongly affected by their non-collisional behaviour during cluster mergers, LTM models of the mass distribution in clusters are by default unable to account for the gravitational effects of collisional mass components, such as the viscous intra-cluster medium (ICM). Although, for massive clusters, the ICM contributes typically only 10 to 15\% of the total cluster mass \citep{2006ApJ...640..691V,2013ApJ...778...14G}, the impact of this fundamental limitation on the results obtained with \texttt{AStroLens} (and other LTM analyses), while small in general, depends on the relaxation state of the clusters under investigation. These effects could be quantified for individual clusters through dedicated X-ray observations in conjunction with detailed lens modeling.
\item[\textbf{Projection effects}] Lastly, a fundamental limitation of \texttt{AStro\-Lens} (and to some extent most strong-lensing analyses) is the assumption that the deflection field is created by a single lens. Since, in reality, the deflection is the cumulative result of all mass concentrations along the line of sight, our approach leads to a  systematically underestimated lensing strength in fields where multiple clusters are superimposed along our line of sight or connected to line-of-sight filaments.  Conversely, fore- or background galaxies from the field (i.e., objects that are not associated with mass overdensities) erroneously selected as likely cluster members can cause the lensing strength to be overestimated.
\end{description}

A more quantitative assessment of the impact of all of the effects mentioned above is  presented in Ebeling et al. (in preparation) which compares \texttt{AStroLens} results derived from ground- and space-based imaging of the same cluster sample.

\section{Conclusion} \label{sec:conclusion}

We describe and present \texttt{AStroLens}, a newly developed LTM algorithm designed to predict the deflection field of massive galaxy clusters solely from the distribution of light emitted by the cluster member galaxies. Building on the principles established and applied successfully in previous work, \texttt{AStroLens} contains a number of new elements. Specifically, the code accounts for the intrinsic alignment between the large-scale mass distribution and the BCG and incorporates several new modeling features, including PIEMD profiles for both galaxy- and cluster-sized mass halos and increased weighting of the BCG. 

We calibrate \texttt{AStroLens}' three global parameters (the core radius of the large-scale mass component, the effective velocity dispersion of the galaxy-scale component, and the scaling coefficient for the additional weight assigned to BCGs) in a $\chi^2$-based minimization procedure anchored by 12 securely identified strong-lensing features in ten galaxy clusters discovered by the eMACS project. The RMS orthogonal distance between the critical lines established by observed strong-lensing features and the critical lines determined by \texttt{AStroLens} for the best-fit parameter set is 2.1\arcsec. Supporting this encouraging performance of our algorithm, tests of the predictive power of 
\texttt{AStroLens} based on a comparison with state-of-the-art lens models from the literature suggest an accuracy of the derived Einstein radii of about 10-20\%.

We use \texttt{AStroLens} to map the gravitational magnification around the strong-lensing regime of 96 X-ray selected galaxy clusters discovered by the eMACS program, based on optical imaging obtained in shallow \textsl{Gemini-N} observations. Modeling all clusters automatically, we identify and characterize 31 lensing clusters from our sample with Einstein radii $\theta\textsubscript{E} > 20"$, 16 with $\theta\textsubscript{E} > 25"$ and eight with $\theta\textsubscript{E} > 30"$. The strongest eMACS lenses are thus comparable to the clusters selected for the Hubble Frontier Fields survey (which produced the deepest observations of lensing clusters to date). \texttt{AStroLens} also proved helpful by pinpointing clusters with very low lensing strength, thereby flagging potential misidentifications in the X-ray selected eMACS sample.

As a first installment toward the public release of the eMACS sample, we provide fundamental physical characterists of our calibration clusters as well as of the ten eMACS clusters with the largest Einstein radii among the sample under investigation.

In ongoing work (Ebeling et al., in preparation), we apply  \texttt{AStroLens} to \textsl{HST} data and assess the impact of systematic effects originating from the quality and characteristics of the optical images from which the input galaxy catalogues are drawn. Although we consider the results presented in this paper highly encouraging, more extensive tests involving sophisticated lens models of individual clusters are needed to firmly establish the accuracy and reliability of \texttt{AStroLens} results.

\section*{Data Availability}
The primary data underlying this article are available in the Gemini Observatory Archive at \hyperlink{https://archive.gemini.edu/searchform}{https://archive.gemini.edu/searchform}, and can be accessed with program identifiers listed in Table \ref{tab:log_obs} under the column ``GMOS-N Observation ID." HST imaging used in this work is available at \hyperlink{https://archive.stsci.edu/access-mast-data}{https://archive.stsci.edu/access-mast-data}, and can be accessed using the SNAPshot program identifiers listed at the beginning of Section \ref{obs}.

\section*{Acknowledgements}

We thank Adi Zitrin for sharing both philosophical and practical insights regarding the computational aspects of his 2012 LTM algorithm that helped us greatly to develop the initial prototype of \texttt{AStroLens}. Johan Richard's extensive strong-lensing expertise was invaluable for the secure identification of strong-lensing features used to calibrate \texttt{AStroLens}' three global modeling parameters. The authors gratefully acknowledge financial support from STScI grants GO-13671, GO-14098, GO-15132, GO-15466, and GO-15843.

%===================================================================================================
\bibliography{lukas_bib.bib}
\bibliographystyle{mnras}

%===================================================================================================

\bsp	% typesetting comment
\label{lastpage}
\end{document}